\documentclass[%
 reprint,
superscriptaddress,
 amsmath,amssymb,
 aps,
prper,
]{revtex4-2}

\usepackage[table,xcdraw]{xcolor}

\usepackage{physics}
\usepackage{xfrac}
\usepackage{braket}
\usepackage{amsmath}
\usepackage{float} 

\usepackage{graphicx}

\usepackage{xcolor}   
\usepackage{hyperref} 

\definecolor{myblue}{rgb}{0, 0, 1}
\definecolor{mygreen}{rgb}{0, 1, 0}

\hypersetup{
    colorlinks=true,
    linkcolor=myblue,
    citecolor=myblue,
    urlcolor=myblue
}

\usepackage{dcolumn}
\usepackage{bm}

\begin{document}


\title{Undergraduate setup for measuring the Bell inequalities and performing Quantum State Tomography}

\author{Raul Lahoz Sanz}
 \email{rlahozsanz@icc.ub.edu}
\affiliation{%
Departament de Física Quàntica i Astrofísica, \\
Facultat de Física, Universitat de Barcelona (QCommsUB group)
}%
\affiliation{Institut de Ciències del Cosmos (ICCUB), Universitat de Barcelona (UB), c. Martí i Franqués, 1, 08028 Barcelona, Spain}

\author{Lidia Lozano Martín}
\affiliation{Institut de Ciències del Cosmos (ICCUB), Universitat de Barcelona (UB), c. Martí i Franqués, 1, 08028 Barcelona, Spain}
\affiliation{
Department of Applied Physics, Universitat de Barcelona, C/Martí i Franquès 1, 08028, Barcelona, Spain
}

\author{Adrià Brú i Cortés}
\affiliation{Departament d’Enginyeria Electrònica i Biomèdica, Universitat de Barcelona (UB),  c. Martí i Franqués, 1, 08028 Barcelona, Spain}

\affiliation{Institut de Ciències del Cosmos (ICCUB), Universitat de Barcelona (UB), c. Martí i Franqués, 1, 08028 Barcelona, Spain}

\author{Martí Duocastella}
\affiliation{
Department of Applied Physics, Universitat de Barcelona, C/Martí i Franquès 1, 08028, Barcelona, Spain
}
\affiliation{
Institute of Nanoscience and Nanotechnology (IN2UB), Universitat de Barcelona (UB), 08028, Barcelona, Spain
}

\author{Jose M. Gómez Cama}
\affiliation{Departament d’Enginyeria Electrònica i Biomèdica, Universitat de Barcelona (UB),  c. Martí i Franqués, 1, 08028 Barcelona, Spain}

\affiliation{Institut de Ciències del Cosmos (ICCUB), Universitat de Barcelona (UB), c. Martí i Franqués, 1, 08028 Barcelona, Spain}

\affiliation{Institut d'Estudis Espacials de Catalunya (IEEC), Edifici RDIT, Campus UPC, 08860 Castelldefels (Barcelona), Spain}

\author{Bruno Juliá-Díaz}
\affiliation{%
Departament de Física Quàntica i Astrofísica, \\
Facultat de Física, Universitat de Barcelona (QCommsUB group)
}%
\affiliation{Institut de Ciències del Cosmos (ICCUB), Universitat de Barcelona (UB), c. Martí i Franqués, 1, 08028 Barcelona, Spain}

\begin{abstract}
The growth of quantum technologies is attracting the interest of many  students eager to learn concepts such as  quantum entanglement  or quantum superposition. However, the non-intuitive nature of these concepts poses a challenge to understanding them. Here, we present an entangled photon system which can perform a Bell test, 
 i.e. the CHSH inequality, and can obtain the complete tomography of 
 the two-photon state. The proposed setup is versatile, cost-effective and 
 allows for multiple classroom operating modes. We present two 
variants, both facilitating the measurement of Bell 
inequalities and quantum state tomography. Experimental 
results showcase successful manipulation of the quantum state 
of the photons, achieving high-fidelity entangled states 
and significant violations of Bell's inequalities. Our 
setup's simplicity and affordability enhances 
accessibility for less specialized laboratories, allowing 
students to familiarize themselves with quantum physics concepts.
\end{abstract}

\maketitle

\section{Introduction}

Quantum superposition and entanglement are key elements 
in the current developments in quantum technologies~\cite{Acín_2018}. 
However, they are elusive concepts with no classical counterpart, 
making them difficult to understand for undergraduate 
students and non-quantum experts. An important step to close 
this gap is through hands on experimentation. By acquiring and analyzing 
data from a quantum entanglement setup, students can get acquainted with 
quantum mechanical concepts and grasp the non-intuitive nature of 
quantum physics. Still, a comprehensive description of such a system,
which is easily accessible for undergraduate students and suitable to 
generate quality data within the time frame of lab practises 
(a couple of hours), remains difficult to find.

Quantum entanglement, the phenomenon by which two particles
become linked so that the state of one affects the state of 
another, regardless of distance, is central in today's quantum technologies. 
Examples include 
quantum computation, e.g. see Shor's algorithm~\cite{shor}, quantum sensing, e.g. enhancing the LIGO
detecting capability~\cite{ligo}, and quantum communications, 
see the Ekert91 protocol~\cite{ekert91}. Despite its importance, 
quantum entanglement has been controversial since the well 
known EPR paper~\cite{einstein1935can}. There, Einstein, Podolsky 
and Rosen argued that the quantum mechanical description 
of a seemingly simple system composed of two particles was most 
likely incomplete. They introduced the notion of hidden variables, 
which at the time seemed more of a philosophical idea than an 
empirically testable one, that would make the description 
of nature complete. 

The situation changed drastically thanks to Bell's article in 
1964~\cite{bell1964einstein}. There, he found a way 
of experimentally setting bounds to the existence of 
hidden variables. He proposed specific experiments to 
prove quantum mechanical predictions could not be explained 
with hidden variables.  Since then, numerous experiments 
have been conducted to verify his predictions. 
From the pioneering experiment by John F. Clauser and Stuart 
Freedman~\cite{freedman1972experimental} onward, all 
have supported the Copenhagen interpretation, emphasizing the intrinsic 
randomness of nature and ruling out 
the possibility of including hidden variables in the 
theory~\cite{aspect1982experimental, rowe2001experimental, hensen2015loophole}. 
The importance of these results was clearly stated by the 
Nobel prize in Physics in 2022, awarded to Alain Aspect, 
John F. Clauser and Anton Zeilinger ``for experiments with entangled photons, establishing the violation of Bell 
inequalities and pioneering quantum information science''~\cite{nobel2022}.

In the last two decades, numerous efforts have been made 
to render this type of experiment accessible to 
undergraduate students. In the universities where these 
laboratory setups have been implemented, e.g.~\cite{galvez2005interference, thorn2004observing, bista2021demonstration, galvez2007quantum, branning2009low, pearson2010hands}, there has been a notable 
improvement in understanding concepts pertaining to 
quantum physics, along with a considerable higher enthusiasm among 
students for such technologies~\cite{PhysRevPhysEducRes.19.010117, lukishova2022fifteen}. 

The main goal of this paper is to present an experimental 
setup for undergraduate students that allows a thorough study of 
the Bell inequalities. For this purpose, we describe 
the implementation, operation and alignment of two such 
setups, enabling students to build them from scratch. Thus, 
our detailed guidelines offer students a pathway to their 
first hands-on experience with quantum concepts.

We see a clear pedagogical value in our work, which can be at 
least twofold: reproducing the experiment can be 
embedded in the syllabus as a final degree project for one or two 
students. Once built, the proposed setup can serve as an 
advanced quantum system in experimental 
laboratories, in our case it is part of the Advanced Quantum Laboratory of 
the Master in Quantum Science and Technology in Barcelona. 

To provide a concise yet self contained document, we 
describe the essential theoretical formalism needed to 
understand the experiment followed by a detailed 
description of the practical setup implementation. The 
article is organized as follows. First, in Sec.~\ref{sec:theory} 
we present the main theoretical concepts involved, 
including the suitable basis states to describe the 
two-photon states and how to perform operations on 
them. This section also provides a theoretical 
description of the quantum state tomography (QST), necessary 
to fully reconstruct the state (subsection ~\ref{sec:tomo}). 
We then introduce Bell inequalities, particularly the 
CHSH inequality~\cite{clauser1969proposed} 
(subsection~\ref{sec:chsh}). The experimental setups 
are described in Sec.~\ref{sec:setups}, providing 
comprehensive descriptions of the proposed implementation.  Subsections~\ref{ss:photprod} and~\ref{sec:phome} explain 
the production and measurement of photons in the two 
respective schemes. The way to align and run the experiment 
is shown in Sec.~\ref{sec:align} while the experimental 
results are collected in Sec.~\ref{sec:exp}. 
Finally, in Sec.~\ref{sec:conclussion}, we outline the qualities 
of our setups and the results of our experiments. We also discuss 
how this work can help to bring these types of concepts 
and technologies closer to a broader and less specialized audience.

\section{Theory}
\label{sec:theory}

In this section we provide the basic theoretical tools needed to 
understand the proposed experiments.

\subsection{Definition of states and operators}
\label{ss:dso}

As the photon is the quantum system of our experiments, we 
start by defining its quantum state. It can be described 
in several useful basis. The most common, the canonical basis 
$\{ \ket{H}, \ket{V} \}$ is formed by the vectors, 
\begin{equation}
    \ket{H} = \begin{pmatrix}
    1 \\
    0 \\  
    \end{pmatrix}
\hspace{0.8cm}
\rm{and}
\hspace{0.8cm}
    \ket{V} = \begin{pmatrix}
    0 \\
    1 \\  
    \end{pmatrix}.
\end{equation}
A set of of different basis, relevant for the experiment described herein, 
are the Diagonal and Antidiagonal basis,the Right-handed and 
Left-handed basis, and the $\alpha$ rotated basis. Note that, in 
this description, photons consist of a two-level quantum mechanical 
system, known as a qubit in the quantum information community. Expressed in 
terms of $\{ \ket{H}, \ket{V} \}$, these basis can be written as 
\begin{equation}
\{ \ket{D}, \ket{A} \} =
    \left\{ \frac{1}{\sqrt{2}} \begin{pmatrix}
    1 \\
    1 \\  
    \end{pmatrix}
,
    \frac{1}{\sqrt{2}} \begin{pmatrix}
    1 \\
    -1 \\  
    \end{pmatrix}
    \right\},
    \label{DA}
\end{equation}

\begin{equation}
\{ \ket{R}, \ket{L} \} =
    \left\{ \frac{1}{\sqrt{2}} \begin{pmatrix}
    1 \\
    -i \\  
    \end{pmatrix}
,
    \frac{1}{\sqrt{2}} \begin{pmatrix}
    1 \\
    i \\  
    \end{pmatrix}
    \right\},
    \label{RL}
\end{equation}

\begin{equation}
\{ \ket{H_\alpha}, \ket{V_\alpha} \} =
    \left\{ \begin{pmatrix}
    \cos\alpha \\
    \sin\alpha \\  
    \end{pmatrix}
,
    \begin{pmatrix}
    - \sin\alpha \\
    \cos\alpha \\  
    \end{pmatrix}
    \right\}.
    \label{VaHa}
\end{equation}

\begin{figure}[tb]
    \centering
    \includegraphics[width=1\linewidth]{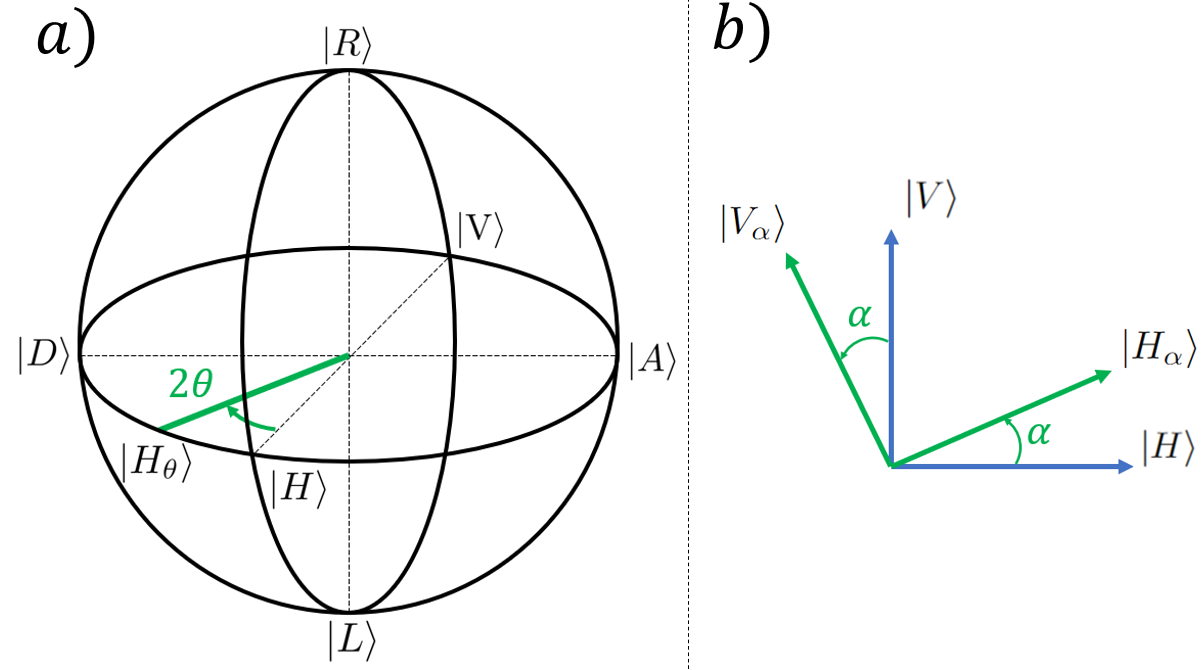}
    \caption{a) Visual representation of the Bloch sphere. All possible states of one single qubit are contained in the surface of the sphere. b) Representation of the $\{ \ket{H}, \ket{V} \}$ and $\{ \ket{H_\alpha}, \ket{V_\alpha} \}$ basis. The equator of the Bloch sphere is the circle generated by the $\{ \ket{H_\alpha}, \ket{V_\alpha} \}$ states.}
    \label{AlphaBasis_BlochSphere}
\end{figure} 

\noindent Where we define the counterclockwise 
direction $\alpha$ as positive when we observe the 
light moving away from us. Thus, in Fig.~\ref{AlphaBasis_BlochSphere} b), the light 
propagates towards the interior of the paper. 

The way to perform unitary operations on 
quantum states without measuring them is using 
retarder plates, particularly half-wave plates (HWP) and 
quarter-wave plates (QWP). As these operations are 
important in our setup, we provide a formal definition 
of their actions in the $\{\ket{H}, \ket{V}\}$ basis. 
The action of a HWP and a QWP with their fast axis 
set at an angle $\theta$  w.r.t the horizontal, i.e. 
the fast axis pointing at the direction $\ket{H_\theta}$, 
is described by
\begin{eqnarray*}
    \mathrm{HWP}_\theta &=& e^{-i\frac{\pi}{2}} \begin{pmatrix}
    \cos2\theta & \sin2\theta \\
    \sin2\theta & - \cos2\theta \\
\end{pmatrix}\,,\nonumber\\
    \mathrm{QWP}_\theta &=& e^{-i\frac{\pi}{4}} \begin{pmatrix}
    \cos^2\theta + i \sin^2\theta & (1-i)\sin\theta\cos\theta \\
    (1-i)\sin\theta\cos\theta & \sin^2\theta + i \cos^2\theta \\
\end{pmatrix}.
\end{eqnarray*}

\noindent Importantly, all these operations can be represented in the Bloch 
sphere, as shown
in Fig.~\ref{AlphaBasis_BlochSphere} a)~\cite{waseem2020quantum}. A HWP (QWP) 
with its fast 
axis set at an angle defined by the vector 
$\ket{H_\theta}$, performs rotations of 
$180^\circ$ ($90^\circ$) of any single photon state 
with respect to the axis defined by the direction 
of the fast axis.

\subsection{Reconstruction of a general two-photon state. 
Quantum State Tomography}
\label{sec:tomo}

The next step is to introduce two-photon states, which 
represent the minimal photonic system which 
can exhibit quantum entanglement. As customary in 
quantum optics and quantum information, we name 
Alice and Bob the two individuals that measure 
the first and the second photon, respectively. 

The general case of non-pure two-photon states can be fully described
by the density matrix. In other words, by experimentally measuring the density matrix, one can
gather all the necessary information to assess two-photon states, a process known as quantum state tomography.
The density matrix can be written as,
\begin{equation}
    \hat{\rho} =
    \begin{pmatrix}
    A_1 & B_1e^{i\phi_1} & B_2e^{i\phi_2} & B_3e^{i\phi_3} \\
    B_1e^{-i\phi_1} & A_2 & B_4e^{i\phi_4} & B_5e^{i\phi_5} \\
    B_2e^{-i\phi_2} & B_4e^{-i\phi_4} & A_3 & B_6e^{i\phi_6} \\
    B_3e^{-i\phi_3} & B_5e^{-i\phi_5} & B_6e^{-i\phi_6} & A_4 \\
    \end{pmatrix}\,,
    \label{eq:dm}
\end{equation}
where the basis used is $\{|HH\rangle, |HV\rangle, 
|VH\rangle, |VV\rangle \}$. 
Note that this matrix is hermitian $\rho=\rho^\dagger$ 
and thus semi-definite positive. Also, the trace has 
to be equal to 1, i.e. $A_1 + A_2 + A_3 + A_4 = 1$. Thus, 
we need $16-1=15$ parameters to fully characterize 
the matrix. For our experiments, it is useful to expand the 
density matrix as a sum of tensor products of two 
Pauli matrices, see for instance~\cite{waseem2020quantum},
\begin{equation}
    \hat{\rho} = \frac{1}{4} \sum_{i,j=0}^{3} S_{ij} \cdot \hat{\sigma_i} \otimes \hat{\sigma_j}\,,
    \label{Tomo}
\end{equation}
 where $\hat{\sigma_i}$ are the identity and the usual Pauli matrices  defined as
\begin{eqnarray*}
    \hat{\sigma_0} = \ketbra{V}{V} + \ketbra{H}{H} = \begin{pmatrix}
    1 & 0 \\
    0 & 1 \\
    \end{pmatrix}\,, \nonumber\\
    \hat{\sigma_1} = \ketbra{D}{D} - \ketbra{A}{A} = \begin{pmatrix}
    0 & 1 \\
    1 & 0 \\
    \end{pmatrix}\,,\nonumber\\
    \hat{\sigma_2} = \ketbra{L}{L} - \ketbra{R}{R} = \begin{pmatrix}
    0 & -i \\
    i & 0 \\
    \end{pmatrix}\,,\nonumber\\
    \hat{\sigma_3} = \ketbra{H}{H} - \ketbra{V}{V} = \begin{pmatrix}
    1 & 0 \\
    0 & -1 \\
    \end{pmatrix}\,.
\end{eqnarray*}
Importantly, the coefficients defining the state, $S_{ij}$ in 
Eq.~(\ref{Tomo}), named Stokes coefficients, can be 
obtained from combined experimental measurements of the two photons 
in the state. For instance, 
\begin{equation}
 S_{00} = P_{\ket{HH}} + P_{\ket{HV}} + P_{\ket{VH}} + P_{\ket{VV}},
\end{equation}
where  $P_{\ket{\sigma \sigma'}}$ is the joint probability 
that Alice and Bob have of obtaining their respective photons 
in the states $\ket{\sigma}$ and $\ket{\sigma'}$ when Alice 
is measuring in the basis  $\{ \ket{\sigma}, \ket{\sigma^\perp} \}$ 
and Bob in the basis $\{ \ket{\sigma'}, \ket{\sigma'^\perp} \}$. The explicit expressions for all Stokes coefficients 
can be found in Appendix~\ref{sec:app-stokes}. This forms the basis
of quantum state tomography.

In our experiment we produce two-photon states which 
are pure states. They are a
 particular case of the general one in Eq.~(\ref{eq:dm}), and can be written as
\begin{equation}
    \ket{\Psi} = a_0 \ket{HH} + a_1 \ket{HV} + a_2 \ket{VH} + a_3 \ket{VV},
    \label{GeneralPureState}
\end{equation}
with $a_i$ ($i=0, 1, 2, 3$) complex coefficients such that $\sum_{i=0}^{3}|a_i|^2 = 1$. 

To compare how similar are two distinct two-photon 
states, $\hat{\rho_1}$ and $\hat{\rho_2}$, we define the fidelity, ${\rm F}(\hat{\rho_1}, \hat{\rho_2})$,~\cite{jozsa1994fidelity},
\begin{equation}
    \rm{F}(\hat{\rho_1}, \hat{\rho_2}) = \left( Tr \left[ \sqrt{\sqrt{\hat{\rho_1}}\hat{\rho_2}\sqrt{\hat{\rho_1}}} \right] \right)^2.
    \label{F}
\end{equation}
Furthermore, if one of the two states under comparison is pure ($\hat{\rho_2} = \ketbra{\Psi_2}{\Psi_2}$), the expression defined 
in Eq.~(\ref{F}) becomes
\begin{equation}
    \rm{F}(\hat{\rho_1}, \hat{\rho_2}) = Tr(\hat{\rho_1}\ketbra{\Psi_2}{\Psi_2}) = \bra{\Psi_2}\hat{\rho_1}\ket{\Psi_2}  \,.
\end{equation}

\noindent The values of the fidelity fall between 
0 and 1. Fidelity 1 is only achieved if both states 
are equal, while fidelity 0 is obtained for orthogonal 
states.

\subsection{Entangled States}
\label{sec:entstate}

A key concept for this work is that of quantum entanglement. 
Working with pure states of the form of 
Eq.~(\ref{GeneralPureState}), a two-photon 
state is said to be entangled if it cannot 
be written as a separable state, 
\begin{equation}
    \ket{\Psi}_{\mathrm{Separable}} = \ket{\psi} \otimes \ket{\varphi}\,. 
\end{equation}
with $\ket{\psi}$ and $\ket{\varphi}$ single photon 
states. 
Note that in a separable state, the outcomes of the measurements 
of Alice and Bob are completely independent, while in 
entangled states, quantum correlations arise between 
the two outcomes. 
 
A set of well-known and useful entangled states are the 
so-called Bell states, 
\begin{align*}
 \ket{\Phi^+} &= {1\over\sqrt{2}} (\ket{HH} + \ket{VV}), \nonumber\\
 \ket{\Phi^- }&= {1\over\sqrt{2}} (\ket{HH} - \ket{VV}), \nonumber\\
  \ket{\Psi^+} &= {1\over\sqrt{2}} (\ket{HV} + \ket{VH}), \nonumber\\
  \ket{\Psi^-} &= {1\over\sqrt{2}} (\ket{HV} - \ket{VH}).
\end{align*}
\noindent These states, in turn, form a basis of the 
two-photon Hilbert space. 
 
\subsection{Bell Test - CHSH inequality}
\label{sec:chsh}

The correlations stemming from the non-separability of states
rose important criticism. Notably, in the so-called 
EPR paradox stated in Ref.~\cite{einstein1935can}, 
it was argued that the 
description of nature is probably incomplete, calling for 
the existence of so-called hidden variables. 
John Bell~\cite{bell1964einstein} introduced the first 
empirical approach to distinguish predictions from 
hidden variable theories. Since then, a series 
of Bell-type inequalities, i.e. Bell tests, have been developed
to check if the quantum state associated to two particles 
follows a non-local behaviour. In particular, quantum mechanics  
produces predictions which violate Bell inequalities, 
i.e. they are incompatible with hidden 
variable theories. In our case, we consider the CHSH 
inequality~\cite{clauser1969proposed}, which was the one 
used by the pioneering article of Aspect and 
collaborators~\cite{aspect1982experimental}, 
and also in the pedagogical setup 
of Ref.-\cite{dehlinger2002entangled_1}. 

\begin{figure}[tb]
    \centering
    \includegraphics[width=1\linewidth]{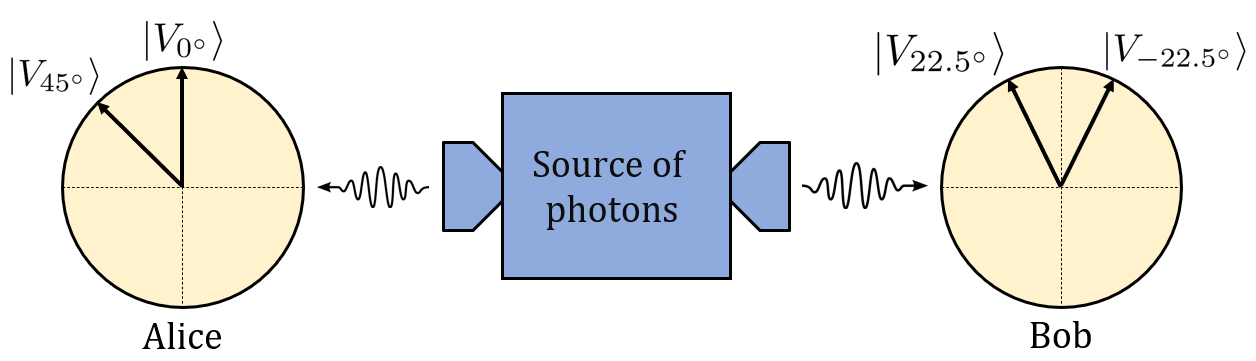}
    \caption{Sketch of the CHSH protocol: A source generates 
    photon pairs always in the same state, sending one to Alice 
    and the other to Bob. They can determine if the received pairs 
    are entangled performing measurements in different bases and sharing their results.}
    \label{CHSH}
\end{figure}

To carry out the Bell test, we consider the scenario 
described in Fig.~\ref{CHSH}. There, a source generates 
pairs of photons, named signal and idler, that are always produced in the same manner, and thus 
in the same quantum state, and that are sent in 
different directions. The receivers of these photons, again 
Alice and Bob, can determine whether 
the pairs of photons they share are entangled or not by 
performing measurements of the individual photons separately and 
communicating the results.

We define the functions $\rm{a}(\alpha)$ ($\rm{b}(\beta)$) as  
$\rm{a}(\alpha) = 1$ ($\rm{b}(\beta) = 1$) if Alice (Bob) 
measures the signal (idler) photon in the state 
$\ket{V_\alpha}$ and $\rm{a}(\alpha) = -1$ ($\rm{b}(\beta) = -1$) 
if Alice (Bob) measures the signal (idler) photon in the 
state $\ket{H_\alpha}$. Then, we define the correlation 
function $E(\alpha, \beta)$, i.e. the average of the product of 
both measurements, as 
\begin{equation}
\begin{split}
    &E(\alpha, \beta) = \langle \rm{a}(\alpha) \cdot \rm{b}(\beta) \rangle = \\
    &= P_{\ket{V_\alpha V_\beta}} - P_{\ket{V_\alpha H_\beta}}  - P_{\ket{H_\alpha V_\beta}}  + P_{\ket{H_\alpha H_\beta}}\,.
\end{split}
\label{E}
\end{equation}

\noindent In the CHSH inequality, 
Alice (Bob) measures the state of the photons in two 
different states, $\alpha = 0^\circ$ ($\beta = 22.5^\circ$) and 
$\alpha' = 45^\circ$ ($\beta' = -22.5^\circ$). 
Thus, Alice and Bob obtain four different values of 
Eq.~(\ref{E}), one for each combination of angles. 
$E(\alpha, \beta)$, $E(\alpha', \beta)$, $E(\alpha, \beta')$ 
and $E(\alpha, \beta')$. Using these four values, we define the 
functions $S$ and $S'$ as,
\begin{equation}
    S = E(\alpha, \beta) + E(\alpha, \beta') + E(\alpha', \beta) - E(\alpha', \beta'),
    \label{CHSH_S}
\end{equation}
\begin{equation}
    S' = E(\alpha, \beta) + E(\alpha, \beta') - E(\alpha', \beta) + E(\alpha', \beta')  \,.
    \label{CHSH_S'}
\end{equation}
These functions are constructed to always 
yield a value between $-2$ and $+2$ when working with classical 
correlations, including the case of variables hidden theories. In contrast, their value falls between 
$-2\sqrt{2}$ and $+2\sqrt{2}$ when we compute the averages with 
quantum mechanics. Specifically, for each Bell 
state, one of them yields a result of zero, while the other provides 
a value equal to $-2\sqrt{2}$ or $+2\sqrt{2}$:
\begin{itemize}
    \item If the two photons are in the state $\ket{\Phi^{+(-)}}$ we obtain $\langle S \rangle = 2\sqrt{2}(0)$  and $\langle S' \rangle = 0(2\sqrt{2})$.
    \item If the two photons are in the state $\ket{\Psi^{+(-)}}$ we obtain $\langle S \rangle = 0(-2\sqrt{2})$ and $\langle S' \rangle = -2\sqrt{2}(0)$.
\end{itemize}

The fact that different Bell states require different Bell 
test functions $S$ and $S'$ is often not emphasized. 
In our case, we can precisely 
control the relative phase between components in 
the wave function of photon pairs along with the use of a 
HWP in one of the photons' paths. 
Thus, with our setups, we have the ability to 
generate the four maximally entangled Bell states, 
as is described later in Sec.~\ref{ss:photprod}. This two 
features set our work apart from other 
pedagogical setups.

\section{Experimental Setups}
\label{sec:setups}

The two setups presented herein consist of a photon pair production 
part followed by a photon detection part. They both enable performing a full two-photon 
state tomography and a Bell test, and 
incorporate improvements at both the technical and 
conceptual level with respect to previous works. Among 
them, our setups allow us to prepare different Bell states, 
which emphasizes the fact that Bell tests are tailored for 
specific states. Also, both options feature a significantly 
simpler optical alignment of the elements of the setup and are fairly robust. 
Importantly, the measurement time is very reasonable: a 
Bell test can be performed in less than one hour, providing 
ample options for lab experimentation at both the graduate 
and master level.  

\begin{figure}[b]
    \centering
    \includegraphics[width=1\linewidth]{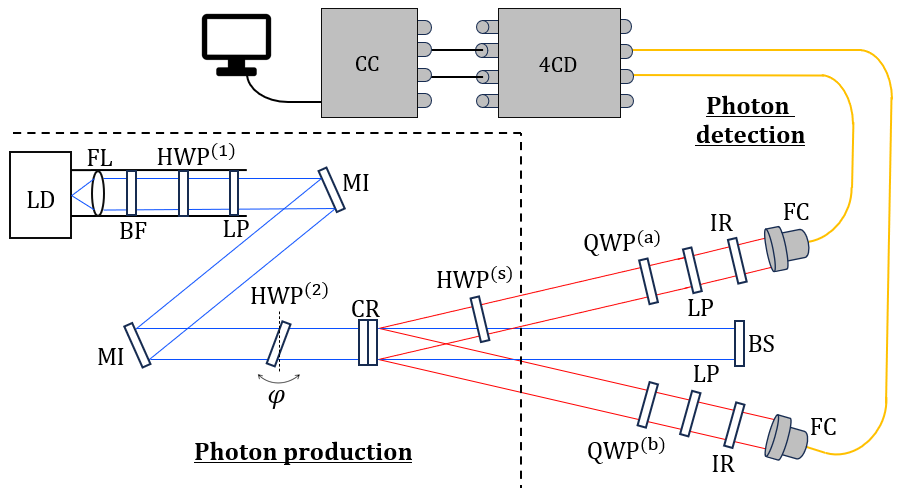}
    \caption{Scheme setup 1. LD. laser diode, FL. focusing lens, BF. blue filter, LP. linear polarizer, MI. mirror, CR. BBO crystals, HWP. half-wave plate, QWP. quarter-wave plate, IR. infrared filter, FC. fiber-coupler lenses, 4CD. 4-channel detector, CC. Coincidence Circuit.}
    
    \label{Setup1}
\end{figure}

\begin{figure}[tb]
    \centering
    \includegraphics[width=1\linewidth]{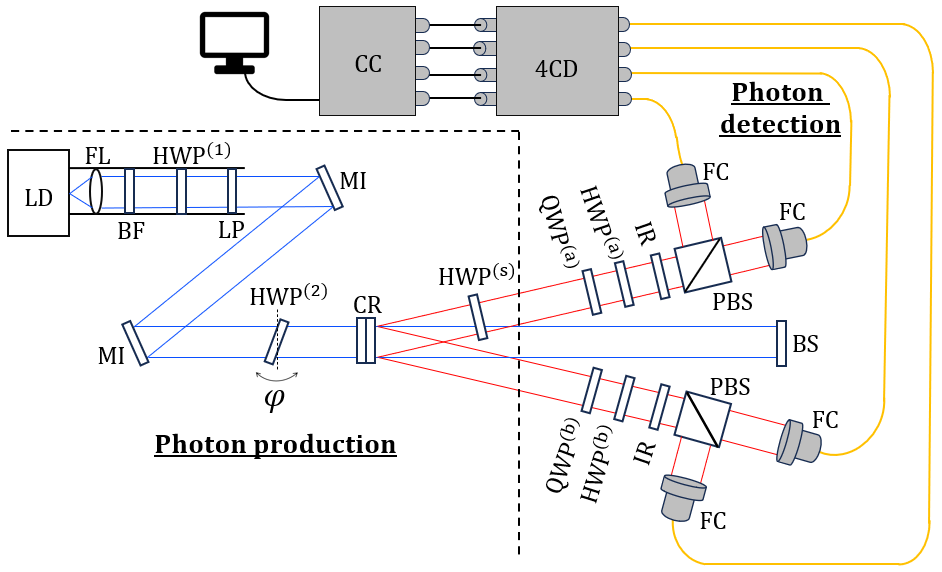}
    \caption{Scheme setup 2.  PBS. polarizing beam splitter. The rest of the elements are labeled in the same way as in Fig.~\ref{Setup1}.}
    \label{Setup2}
\end{figure}

A detailed list of the necessary 
equipment for each setup is compiled in 
Appendix~\ref{sec:app1}. The main difference between the two setups
lies in the photon detection process. The first setup, illustrated in Fig.~\ref{Setup1}, employs only 
two inputs of the 4-channel detector \textsf{(SPCM-AQ4C, Excelitas Technologies)} and measures the polarization of the light 
using a QWP and a polarizer. While simpler in terms of optical elements, this option is slower for measurements. 
It can only provide the number of coincident photon counts passing through both polarizers in a single measurement. That is, pairs of photons that pass through both polarizers without being stopped and are detected simultaneously. 
The second setup, depicted in Fig.~\ref{Setup2}, needs 
all 4 inputs of the detector and directs photons to different 
detectors based on their polarization using polarizing beam 
splitters (PBS). Although this option requires more optical 
elements, it is faster for measurements as it allows for the 
measurement of photon counts in any of the states of a given 
basis in a single measurement.

In both of our setups, as in some other pedagogical 
experiments~\cite{thorn2004observing, bista2021demonstration, pearson2010hands}, collimating lenses \textsf{(F810FC-780, ThorLabs)} and optical fibers are used to capture photons. 
This offers a significant advantage in alignment as is 
discussed in Sect.~\ref{ss:aligFC}. In some previous 
works~\cite{dehlinger2002entangled_1, dehlinger2002entangled_2, galvez2005interference} the alignment of the system and 
the capture of photons was performed without the use of 
optical fibers. To count coincidences between two channels we have 
replicated the circuit described in Ref.~\cite{dehlinger2002entangled_2} modifying the capacitors to reduce the coincidence window to 90~ns and adding USB connectivity. 

\subsection{Photon Production}
\label{ss:photprod}

As shown in  Fig.~\ref{Setup1} and Fig.~\ref{Setup2}, the two-photon production part is similar in both setups. It also shares many 
elements with previous works, in particular with that 
of Dehlinger and 
Mitchell~\cite{dehlinger2002entangled_1,dehlinger2002entangled_2}. In more detail, we use a 405nm 
laser beam \textsf{(L404P400M, ThorLabs)} working at 400mW 
that emits light horizontally-polarized
\begin{equation*}
    \ket{\Psi} = \ket{H}_{\rm{Pump}} \,.
\end{equation*}
To switch from horizontal to diagonal 
light with almost no energy loss, we employ a HWP with its optical axis set at an angle 
$\theta = 22.5^\circ$ 
\begin{equation*}
    \mathrm{HWP}^{(1)}_{\theta = 22.5^\circ} \ket{H} = e^{i \cdot \frac{\pi}{2}} \ket{D} = \ket{D}.
\end{equation*}
Additonally, we place a polarizer set at an angle of $45^\circ$ to further ensure the desired polarization sate. Thus, the quantum state after the polarizer reads,
\begin{equation*}
    \ket{\Psi} = \ket{D}_{\rm{Pump}}\,.
\end{equation*}
Afterwards, the light gets reflected by the two 3D-precision 
mirrors and passes through a HWP with its fast axis parallel to 
the optical table. This HWP is mounted on a goniometer \textsf{(RP01/M, ThorLabs)} that allows us to tilt it 
around the axis perpendicular to the optical table. With 
this tilt angle 
$\varphi$ we can vary the relative phase ($\phi(\varphi)$) 
between the $\ket{H}$ and $\ket{V}$ component.
\begin{equation}
    \ket{\Psi} = \frac{1}{\sqrt{2}}\left( \ket{H}_{\rm{Pump}} + e^{i \cdot \phi(\varphi)} \ket{V}_{\rm{Pump}}
    \right)\,.
    \label{Psi_Pump}
\end{equation}

Note that, at this stage, we have produced a photon 
in a superposition of both horizontal and vertical polarization. To generate entangled photons, we 
exploit a phenomenon called 
spontaneous parametric down-conversion 
(SPDC) Ref.~\cite{dehlinger2002entangled_2}. To this end, we place a pair of barium borate (BBO) crystals (both Type I, 
cut at a phase matching angle $\theta=29.2^\circ$, with 
dimensions $6\times6\times0.1$mm, optically contacted on and 
each one rotated 90$^\circ$ with respect to the other) in the light path.  Upon interaction with the BBO crystals, 
an initial single photon, called pump, can generate two down-converted (and thus, less energetic) photons. Although the probability 
of this process is low, one in a million, the high photon flux that reaches the BBO crystal ensures
repeatable generation of Bell states. 

\begin{figure}[t]
    \centering
    \includegraphics[width=1\linewidth]{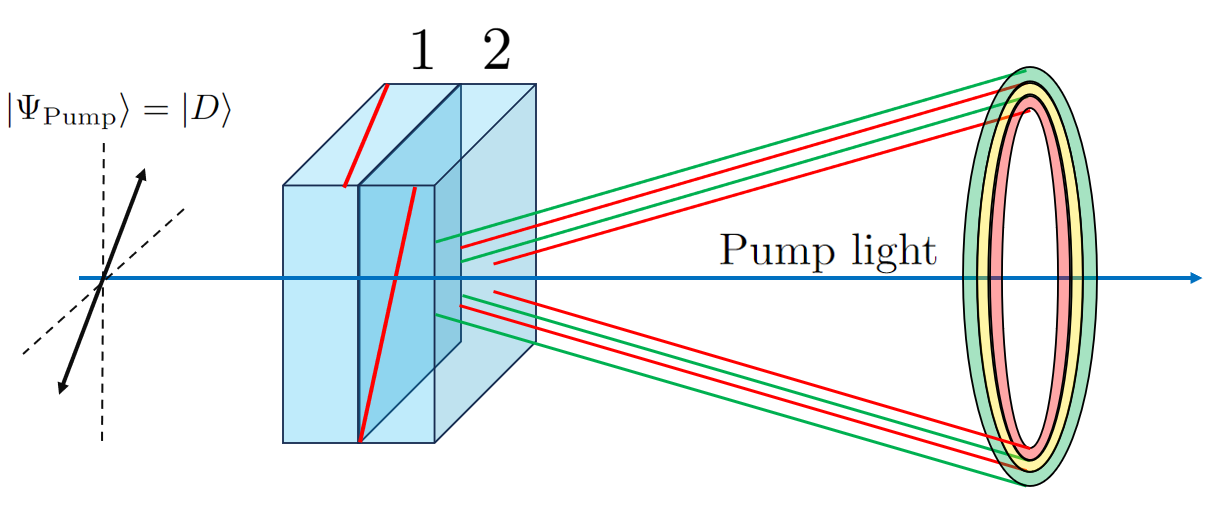}
    \caption{Scheme of the production of entangled photons using two Type-I BBO crystals. The optical axes of the crystals represented as red line are forming $90^\circ$.}
    \label{PairBBO}
\end{figure}

The plane formed by the optical axis of the 
BBO crystal and the direction of propagation of the incident 
pump photon is known as the SPDC plane. Only a pump photon with polarization contained 
in the SPDC plane can experience SPDC and generate two photons. In this case, both photons feature a perpendicular polarization with respect to that of the incident pump photon. Instead, if the polarization of the pump photon 
is perpendicular to the BBO plane, the BBO crystal does not produce 
pairs of photons~\cite{kwiat1999ultrabright}. With this consideration, by illuminating the first 
(second) BBO crystal with horizontal (vertical) light, 
pairs of photons can be produced, both with vertical 
(horizontal) polarization, as shown in Fig.~\ref{PairBBO}. The green and red cones 
are the $V$-polarized and $H$-polarized light cones, 
respectively. 

As the photons produced in the first BBO crystal 
have extraordinary polarization in the second BBO crystal, then, a 
relative phase $\phi_{\rm{BBO}}$ appears between the pair of 
photons produced in the first and the second crystal.

\begin{equation}
    \ket{H}_{\rm{Pump}} \xrightarrow{\rm BBO's} \ket{V}_{\rm{s}}\otimes\ket{V}_{\rm{i}}
    \label{SPDC_1},
\end{equation}
\begin{equation}
    \ket{V}_{\rm{Pump}} \xrightarrow{\rm BBO's} e^{i \cdot \phi_{\rm{BBO}}} \cdot \ket{H}_{\rm{s}}\otimes\ket{H}_{\rm{i}}.
    \label{SPDC_2}
\end{equation}

\noindent In our experiments, we shine the BBO crystals with a diagonally-polarized light, that is, light in an equally superposition between the $\ket{H}$ and $\ket{V}$ states. These photons can be down-converted in both crystals. Thus, in the region of space 
where both light cones overlap (yellow region in Fig.~\ref{PairBBO}) the photons that we receive are indistinguishable, i.e.  
we cannot tell in which BBO crystal they were generated. 
What we do know is that, if we measure the polarization of one 
of them, the polarization of the other is the same. It is precisely this indistinguishability between two-photon paths gives rise to the entanglement.

Let us consider the pair of BBO crystals with their optical axes pointing in the vertical and horizontal direction. When one pump photon in the state Eq.~(\ref{Psi_Pump}) goes through them and suffers SPDC, following Eqs.~(\ref{SPDC_1}) and~(\ref{SPDC_2}) produces

\begin{equation}
    \ket{\Psi_{\rm{EPR}}} = \frac{1}{\sqrt{2}}\left( \ket{VV} + e^{i \cdot (\phi(\varphi)+\phi_{\rm{BBO}})} \ket{HH}
    \right)\,.
    \label{Psi_EPR}
\end{equation}

\noindent Changing the tilt angle $\varphi$ of the 
$\rm{HWP}^{(2)}$ we can control the relative phase 
between $\ket{VV}$ and $\ket{HH}$ components. Let us 
call $\varphi^+$ and $\varphi^-$ the angles for which 
we obtain
\begin{equation}
    \begin{cases}
        \varphi = \varphi^{+} \hspace{0.3cm}\xrightarrow{}\hspace{0.3cm}
        e^{i \cdot (\phi(\varphi)+\phi_{\rm{BBO}})} = 1
        \\
        \varphi = \varphi^{-} \hspace{0.3cm}\xrightarrow{}\hspace{0.3cm}
        e^{i \cdot (\phi(\varphi)+\phi_{\rm{BBO}})} = -1.
    \end{cases}    
\end{equation}

Finally, the four Bell states can be produced with this 
setup by adding a HWP. In particular, a HWP ($\rm{HWP}^{(s)}$) placed 
in the optical path corresponding to the signal photons, 
enables obtaining all four Bell states. As shown in Table~\ref{BellStates}, these states depend on the angles 
$\varphi$ and $\theta_s$, where $\theta_s$ 
is the angle that forms the fast axis of the $\rm{HWP}^{(s)}$ 
with respect to the horizontal direction. 

\begin{table}[t]
\begin{ruledtabular}
\begin{tabular}{ccc}
\textrm{Bell State}&
\textrm{Angle} $\theta_s$ &
\multicolumn{1}{c}{\textrm{Angle} $\varphi$}\\
\colrule
$\ket{\Phi^{+}} = \frac{1}{\sqrt{2}} \left( \ket{HH} + \ket{VV} \right)$ & $0^\circ$ & $\varphi = \varphi^{+}$ \\ 
\hline

$\ket{\Phi^{-}} = \frac{1}{\sqrt{2}} \left( \ket{HH} - \ket{VV} \right)$ & $0^\circ$ & $\varphi = \varphi^{-}$ \\ 
\hline

$\ket{\Psi^{+}} = \frac{1}{\sqrt{2}} \left( \ket{HV} + \ket{VH} \right)$ & $45^\circ$ & $\varphi = \varphi^{+}$ \\ 
\hline

$\ket{\Psi^{-}} = \frac{1}{\sqrt{2}} \left( \ket{HV} - \ket{VH} \right)$ & $45^\circ$ & $\varphi = \varphi^{-}$ \\
\end{tabular}
\end{ruledtabular}
\label{BellStates}
\caption{Value of the variables $\varphi$ and $\theta_s$ for the production of all four Bell States.}
\end{table}

\subsection{Photon detection}
\label{sec:phome}

The key distinction between both setups lies in the photon 
detection part. In particular, in the following three aspects: 1) the number of detector channels employed, 2) the way we perform unitary transformations on 
the photon individually and, 3) the way the polarization 
is measured.

The state of one photon can be expressed 
in any of the basis introduced previously, 
\begin{equation}
    \begin{split}
        \ket{\Psi} &= C_{V} \ket{V} + C_{H} \ket{H} \\
        &= C_{V_\alpha} \ket{V_\alpha} + C_{H_\alpha} \ket{H_\alpha} \\
        &= C_{R} \ket{R} + C_{L} \ket{L}, \\
    \end{split}  
\label{?_State}
\end{equation}

\begin{figure}[t]
    \centering
    \includegraphics[width=1\linewidth]{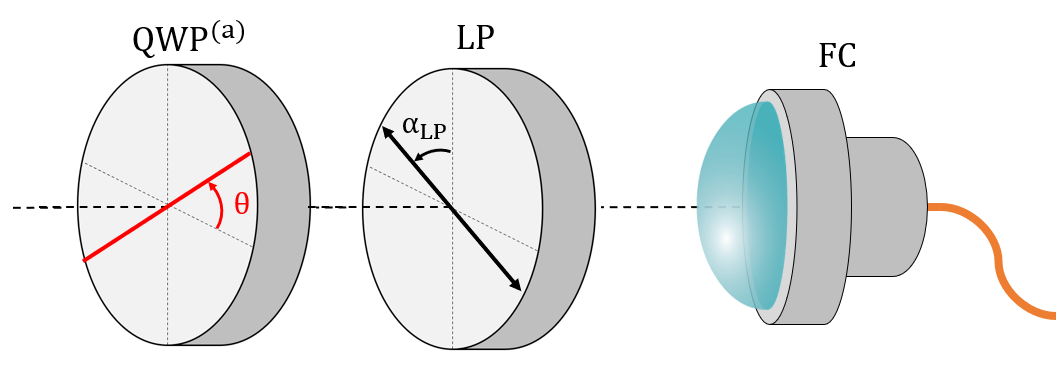}
    \caption{Measurement scheme in the signal photons' arm in Setup 1. Photons travel from left to right, first passing through the QWP with its fast axis (red line) set at an angle $\theta$ with respect to the horizontal. They then go through the linear polarizer, oriented at an angle $\alpha_{\mathrm{LP}}$ relative to the vertical. In the idler photons' arm, the scheme is the same.}
    \label{MeasSetup1}
\end{figure}

\noindent where $C_i$ are complex numbers. The squared modulus of these coefficients represents the probability of finding the photon in that state. Our main objective is to measure the number of photons that reach our detectors in each state of a given basis. However, we are limited in the information that we can gather. For example, by varying the polarizer angle $\alpha$ with respect to the vertical direction in the setup shown in Fig.~\ref{Setup1}, we are restricted to measure the states located on the equator of the Bloch sphere $\{ \ket{V_\alpha}, \ket{H_\alpha} \}$ (Fig.~\ref{AlphaBasis_BlochSphere} a). The setup depicted in Fig.~\ref{Setup2} is even more restrictive, allowing access to only the basis $\{ \ket{V}, \ket{H} \}$. To measure photons in the various bases of interest in each setup, which is needed for the Bell test and quantum state tomography experiments, retarder plates are required. A brief guideline on how to use them is presented in the following subsections.

\subsubsection{Measurements in setup 1}
\label{ss:ms1}

In this case, the way we have to measure the photons polarization is by using first a QWP and then a linear polarizer, as depicted in Fig.~\ref{MeasSetup1}. If we place the QWP at $\theta = \alpha$,
    
    \begin{equation*}
    \begin{split}
        &\rm{QWP}^{(a)}_{\theta = \alpha}(\ket{H_{\alpha}}) = \ket{H_{\alpha}}, \\
        &\rm{QWP}^{(a)}_{\theta = \alpha}(\ket{V_{\alpha}}) = \ket{V_{\alpha}},
    \end{split}
    \end{equation*}
    
\noindent the QWP acts as an identity operator for the states $\ket{V_{\alpha}}$ and $\ket{H_{\alpha}}$. Therefore, the QWP can be removed as it does not alter the state. Then we can measure 
the number of photons in the $\ket{V_{\alpha}}$ and $\ket{H_{\alpha}}$ states by simply placing the polarizer at an angle $\alpha_{\mathrm{LP}} = \alpha$ and $\alpha_{\mathrm{LP}} = \alpha + 90^\circ$ respectively. On the other hand, if we place the QWP at $\theta = 45^\circ$,
    \begin{align*}
        &\rm{QWP}^{(a)}_{\theta = 45^\circ}(\ket{L}) = \ket{H}, \\
        &\rm{QWP}^{(a)}_{\theta = 45^\circ}(\ket{R}) = \ket{V}, 
    \end{align*}
    
\noindent the entire $\ket{R}(\ket{L})$ component of our state described in the 
Eq.~(\ref{?_State}) becomes $\ket{V}(\ket{H})$. Therefore, if we place the polarizer at an angle $\alpha_{\mathrm{LP}} = 0^\circ$ ($\alpha_{\mathrm{LP}} = 90^\circ$), we obtain the same statistics as if we could measure the state of our photons before the operations with the retarder plates in the state $\ket{R}$ ($\ket{L}$). Summarizing the procedure in a table:

\begin{table}[H]
\begin{ruledtabular}
\begin{tabular}{ccc}
\textrm{Angle QWP} & \textrm{Angle LP} & \multicolumn{1}{c}{\textrm{Counts detected}}\\
\colrule
 $\alpha$ & $\alpha$ & $N_{\ket{V_\alpha}}$ \\ 
\hline

 $\alpha$ & $\alpha + 90^\circ$ & $N_{\ket{H_\alpha}}$  \\ 
\hline

 $45^\circ$ & $0^\circ$ & $N_{\ket{R}}$  \\ 
\hline

 $45^\circ$ & $90^\circ$ & $N_{\ket{L}}$   \\
\end{tabular}
\end{ruledtabular}
\label{TableMeasSetup2}
\caption{Value of the angles of the QWP and the linear polarizer for obtaining all different photon-state statistics using the setup depicted in Fig.~\ref{Setup1}.}
\end{table}

\subsubsection{Measurements in setup 2}
\label{ss:ms2}

\begin{figure}[t]
    \centering
    \includegraphics[width=1\linewidth]{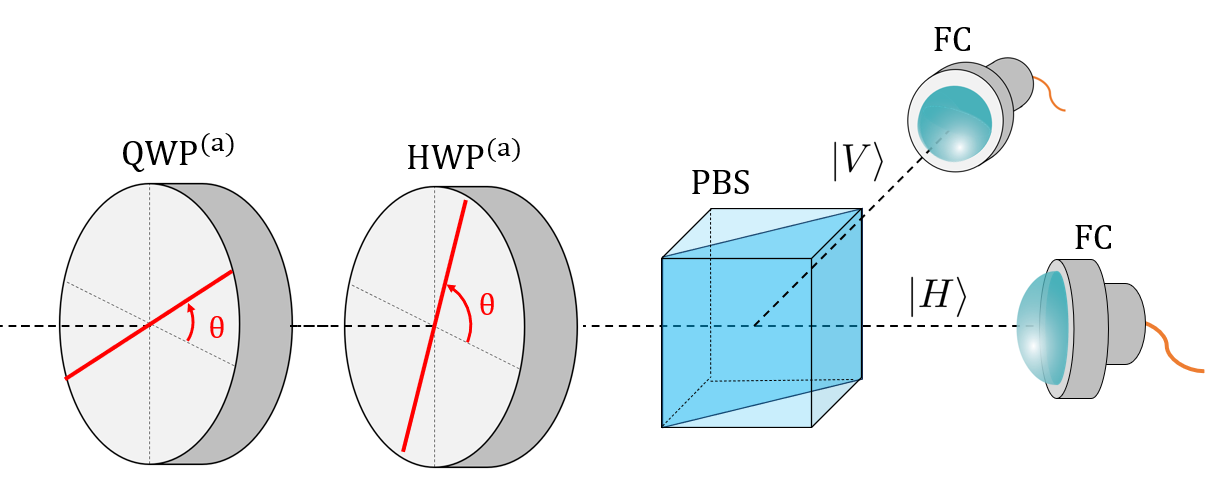}
    \caption{Measurement scheme in the signal photons' arm in Setup 2. Photons travel from left to right, first passing through the QWP and HWP before going through the polarizing beam splitter (PBS). In the PBS, photons with vertical polarization get reflected and photons with horizontal polarization gets transmitted.}
    \label{MeasSetup2}
\end{figure}

In the second setup, we measure the photons polarization using first a QWP, then a HWP and finally a polarizing beam splitter (PBS), as depicted in Fig.~\ref{MeasSetup2}. Thus, photons with vertical (horizontal) polarization are collected by the fiber-coupling lens placed in the reflected (transmitted) path of the PBS. If we place the QWP at $\theta = \alpha$ and the HWP at $\theta = \frac{\alpha}{2}$, we have 
\begin{equation*}
    \begin{split}
        &\rm{HWP}^{(a)}_{\theta = \frac{\alpha}{2}}(\rm{QWP}^{(a)}_{\theta = \alpha}(\ket{H_{\alpha}})) = \rm{HWP}^{(a)}_{\theta = \frac{\alpha}{2}}(\ket{H_{\alpha}}) = \ket{H}, \\
        &\rm{HWP}^{(a)}_{\theta = \frac{\alpha}{2}}(\rm{QWP}^{(a)}_{\theta = \alpha}(\ket{V_{\alpha}})) = \rm{HWP}^{(a)}_{\theta = \frac{\alpha}{2}}(\ket{V_{\alpha}}) = \ket{V},
    \end{split}
\end{equation*}

\noindent where the $\ket{V_\alpha}$ and $\ket{H_\alpha}$ components of any arbitrary state described in the eq. (\ref{?_State}) are transformed into $\ket{V}$ and $\ket{H}$ components respectively. When this photons goes through the PBS, in the reflected (transmitted) path, the vertically (horizontally) polarized photons follows the same statistics as the photons in the state $\ket{V_\alpha}$ ($\ket{H_\alpha}$) before the retarder plates. On the other hand, if we place the QWP at $\theta = 45^\circ$ and the HWP at $\theta = 0^\circ$, we have
\begin{equation*}
    \begin{split}
        &\rm{HWP}^{(a)}_{\theta = 0^\circ}(\rm{QWP}^{(a)}_{\theta = 45^\circ}(\ket{L})) = \rm{HWP}^{(a)}_{\theta = 0^\circ}(\ket{H}) = \ket{H}, \\
        &\rm{HWP}^{(a)}_{\theta = 0^\circ}(\rm{QWP}^{(a)}_{\theta =45^\circ}(\ket{R})) = \rm{HWP}^{(a)}_{\theta = 0^\circ}(\ket{V}) = \ket{V},
    \end{split}
\end{equation*}

\noindent the entire $\ket{R}(\ket{L})$ component of our state becomes $\ket{V}(\ket{H})$. Therefore, in the reflected (transmitted) path of the PBS, the photons follows the same counting statistics as if we will be able to measure the photons in the states$\ket{R}$ and $\ket{L}$ respectively, before the operations with the retarder plates. Summarizing the procedure in a table:

\begin{table}[t]
\begin{ruledtabular}
\begin{tabular}{ccc}
\textrm{Angle QWP} & \textrm{Angle HWP} & \multicolumn{1}{c}{\textrm{Counts detected in the}} \\
\textrm{} & \textrm{} & \multicolumn{1}{c}{\textrm{refelected (transmitted) path}} \\
\colrule
 $\alpha$ & $\sfrac{\alpha}{2}$ & $N_{\ket{V_\alpha}}$ ($N_{\ket{H_\alpha}}$) \\ 
\hline
 $45^\circ$ & $0^\circ$ & $N_{\ket{R}}$ ($N_{\ket{L}}$)  \\ 

\end{tabular}
\end{ruledtabular}
\label{TableMeasSetup2}
\caption{Value of the angles of the QWP and the HWP for obtaining all different photon-state statistics in the reflected and transmitted path of the PBS, using the setup depicted in Fig.~\ref{Setup2}.}
\end{table}

\section{Alignment of the setup}
\label{sec:align}

In this section we explain in full detail how to align our experimental setup. 

\subsection{Alignment of the pump laser and detectors}
\label{ss:aligFC}

The first thing to check once we have all the elements assembled 
is whether the pump laser travels parallel to the optical 
table and through the center of the setup. To do this, we 
rely on the precision mounts (\textsf{KS1, ThorLabs}) of 
the two mirrors on which the pump beam is reflected. 

Subsequently, we need to verify that the fiber-coupling 
lenses \textsf{(F810FC-780, ThorLabs)} are capturing photons 
from the BBO crystals. For this purpose, we introduce light 
through the other output of the optical fiber and ensure 
that all the light spots generated by the fiber-coupling 
lenses points at the BBO crystals using the precision mount (\textsf{KS1, ThorLabs}) in which the fiber-coupling lenses are fixed. Once this is done, we know that the lenses are collecting photons from the BBO crystals.

\subsection{Optimal position for the rail angles}

To increase the number of collected photons, we need to find 
the optimal angles of the detectors rails (and therefore the fiber-coupler lenses) for which we detect the maximum number of coincident counts. We know that we are indeed detecting 
coincident counts from the BBO crystals when these 
counts deviate at least an order of magnitude from the 
counts that would be expected by mere chance, known 
as the accidental counts ($N_{acc}$). These accidental 
counts depend on the number of counts detected individually 
by each of the detectors ($N_a$ and $N_b$), as well as 
the measurement duration (T) and the coincidence window 
($\tau$), according to the following equation
\begin{equation}
    N_{acc} = \frac{N_a \cdot N_b \cdot \tau}{\rm T}.
    \label{N_acc}
\end{equation}

In our case, the value for the coincidence window is fixed an equal to $90$ns. To find the optimal position of the rails, we conduct a 
study in which, for a certain angle of the signal 
rail ($\Theta_s$), we record the number of detected 
coincidences for various positions of the idler rail ($\Theta_i$). 
\begin{figure}
    \centering
    \includegraphics[width=1\linewidth]{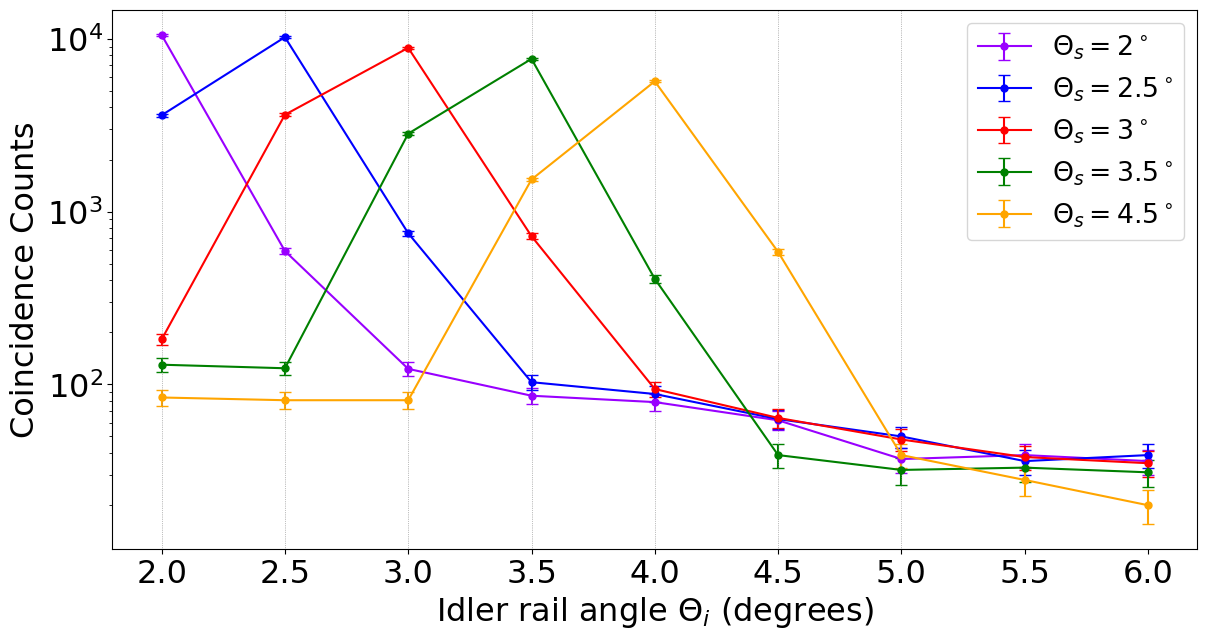}
    \caption{Coincidence counts and averaged accidental counts for different angular positions of the metallic rails. Each of the measurements takes 30 seconds.}
    \label{OptimalRailAngle}
\end{figure}

In Fig.~\ref{OptimalRailAngle}, we find the maximum number 
of coincidence counts for each case when one detector is approximately 
at the same angle of the other detector. In particular, 
we find the maximum number of coincidence counts when 
both detectors are between $2^\circ$ and $2.5^\circ$. 
For mechanical reasons, when assembling all the necessary 
elements to take measurements on the metal rails, we 
cannot position both detectors at less than $2.5^\circ$. 
Therefore, from now on, in the remaining measurements, 
both detectors will be fixed at an angle 
$\Theta_s = \Theta_i = 2.5^\circ$ with respect to the 
pump laser beam. At these angles, the number of coincidence 
counts differs by at least two orders of magnitude compared to 
the number of accidental counts. This indicates that the 
coincidence counts we detect come from photon pairs produced 
in the BBO crystals. It is also worth mentioning that the 
number of dark counts of our detector is around 350 counts 
per channel per second.

\subsection{Optimal position of the BBO crystals. 
Finding the direction of the optical axes}

In general we do not know a priori the direction of the 
optical axes inside each BBO crystal. We know that a BBO 
crystal only produces photons when the polarization of 
the light is contained in the SPDC plane of the 
crystal~\cite{kwiat1999ultrabright}. In our case one 
BBO crystal is rotated $90^\circ$ with respect to the 
other. Thus, if we shine the BBO crystals with pump light 
horizontally polarized and fix them inside a rotation 
mount \textsf{(KS1RS, ThorLabs)}, that allows us to rotate 
both crystals simultaneously, we expect to find four 
angles in which we only detect photons in the state 
$\ket{VV}$. This occurs because, at these angles, one 
BBO crystal aligns its SPDC plane parallel to the pump 
light polarization, generating photon pairs in the state 
$\ket{VV}$ while the other crystal aligns its SPDC 
plane perpendicular to the pump light polarization, 
ceasing photon production.

\begin{figure}[t]
    \centering
    \includegraphics[width=1\linewidth]{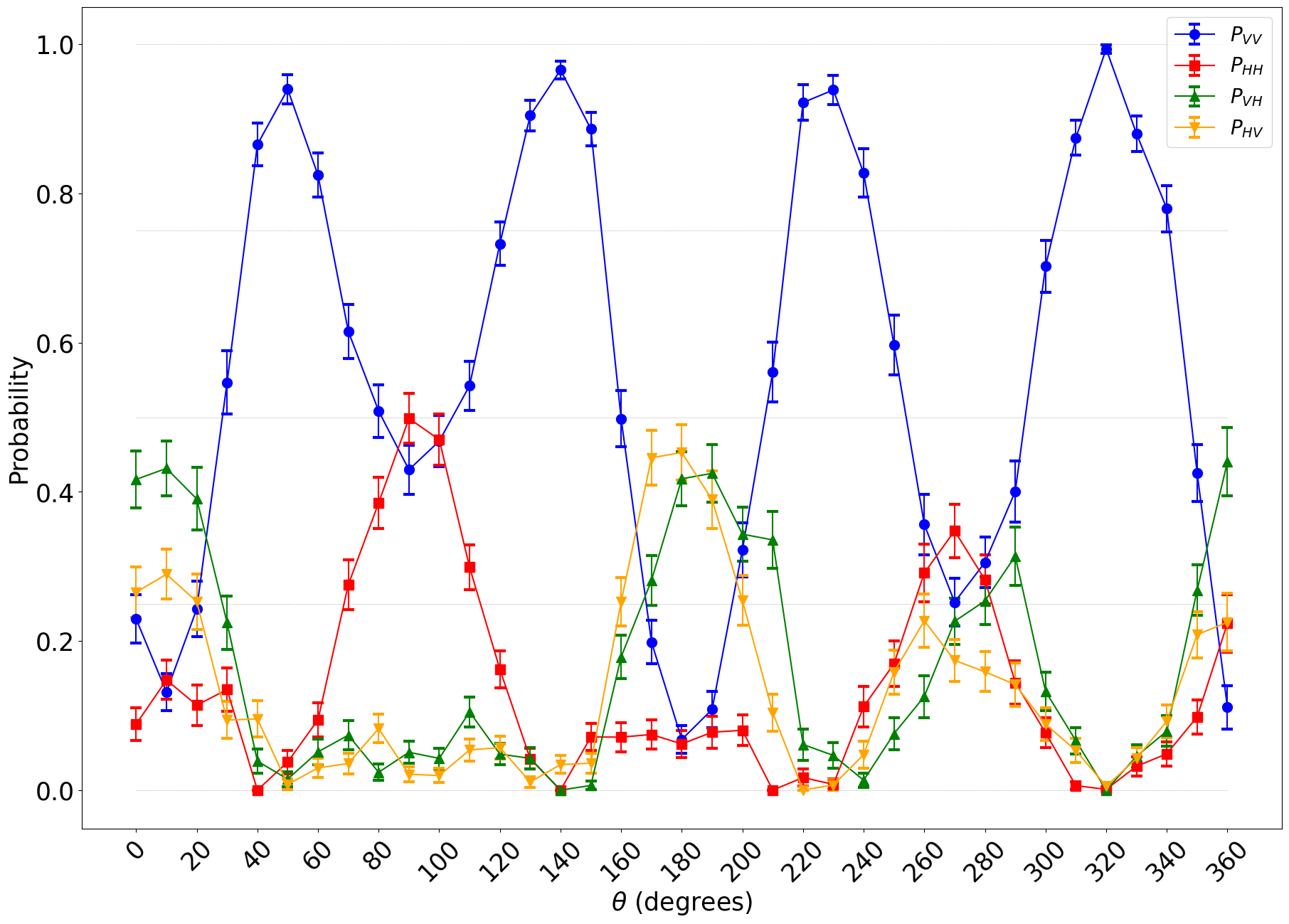}
    \caption{Coincidence probabilities measured for various BBO crystal angles and photon-pair states. Each measurement lasts 30 seconds.}
    \label{OptimalBBOangle}
\end{figure}

Then, we rotate the BBO crystals in steps of $10^\circ$ 
and for each angle we measure the photons in the states 
$\ket{VV}$, $\ket{VH}$, $\ket{HV}$ and $\ket{HH}$, 
obtaining the dependence shown in Fig.~\ref{OptimalBBOangle}.

The maximal signal of photons in the $\ket{VV}$ state is found 
for angles $45^\circ$, $135^\circ$, $225^\circ$ and $315^\circ$, 
see Fig.~\ref{OptimalBBOangle}. These are the optimal angles 
for the production of entangled photons when we shine the 
crystals with diagonally-polarized light, as one of the 
crystals have the optical axis pointing in the $\ket{H}$ 
direction and the other in the $\ket{V}$ direction.

\subsection{Finding the optimal phase matching angle}

Once we have the optical axes of the BBO crystals pointing 
in the $\ket{V}$ and $\ket{H}$ directions, we illuminate them 
again with diagonally polarized pump light. 

Our BBO crystals are cut at a phase-matching angle of 
$29.2^\circ$. That means that the optical axis of the 
crystal forms $29.2^\circ$ with respect to the direction 
of the pump photons. This angle is not optimized for 
the wavelenght that we are working with 
(405nm)~\cite{cheng2022computational}. However, using the 
precision mount in which the crystals are 
placed (\textsf{KS1RS, ThorLabs}), we can slightly tune 
the angle that forms the axis of each BBO crystal with 
respect to the pump light. This optimizes the production of 
photon pairs, as it offers three precision screws, 
arranged in an ``L'' shape. With the screw at the upper 
end, we change the phase-matching angle of the crystal 
whose optical axis is pointing in the vertical direction.  
The screw at the lower end allows us to vary the 
phase-matching angle of the crystal whose optical axis is 
pointing in the horizontal direction.

As we vary the phase-matching angle of the crystal whose 
optical axis is pointing in the vertical (horizontal) 
direction, we measure the photons reaching the detectors 
in the $\ket{HH}$ ($\ket{VV}$) states. This is done 
until we find the optimal phase-matching angle, i.e., 
the angle for which the highest number of photons is detected.

\subsection{Finding the relative 
phase-shift dependence 
with the tilt angle of the HWP$^{(2)}$}

The $\rm{HWP}^{(2)}$ has its fast axis pointing in the 
$\ket{H}$ direction, but is placed in a mount that 
allows to tilt this retarder plate around an axis 
perpendicular to the optical 
table \textsf{(RP01/M, ThorLabs)}. This permits us to 
adjust the relative phase between the $\ket{VV}$ and 
the $\ket{HH}$ photons produced in the BBO crystals, 
as is described in Eq.~(\ref{Psi_EPR}).

\begin{figure}[t]
    \centering
    \includegraphics[width=1\linewidth]{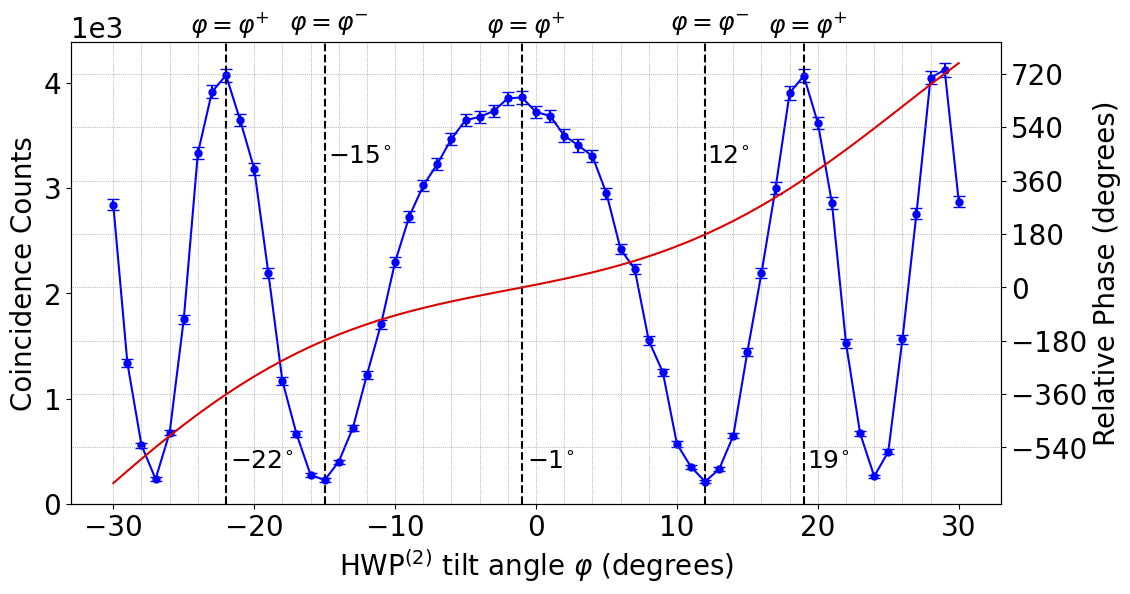}
    \caption{The dependence of coincidence counts, represented by blue dots, and relative phase shift, depicted as a red solid line, on the tilt angle $\varphi$ of the $\rm{HWP}^{(2)}$}
    \label{RelativePhaseDependence}
\end{figure}

If we measure the photons produced by the crystals in the state $\ket{D}_s \otimes \ket{D}_i$ while we vary the tilt angle $\varphi$ of the $\rm{HWP}^{(2)}$, we expect to find the number of coincident counts following the dependence
\begin{equation}
\begin{split}
    N_{\ket{DD}}(\varphi) &\propto |\braket{DD | \Psi_{\rm{EPR}}}|^2 \\
    &\propto \frac{1}{4} \cdot (1+\cos\phi'),
\end{split}
\end{equation}

\noindent where $\ket{\Psi_{\mathrm{EPR}}}$ is the state 
defined in Eq.~(\ref{Psi_EPR}) and 
$\phi' = \phi(\varphi)+\phi_{\rm{BBO}}$. 
When $\varphi = \varphi^{+}+2\pi n$ 
($\varphi = \varphi^{-}+2\pi n$), with $n \in \mathbb{Z}$ 
we expect to find a maximum (minimum) in the number of 
pairs of photons in the state $\ket{DD}$~\cite{kwiat1999ultrabright}.

This dependence of the number of coincidence counts in the state $\ket{DD}$ with the tilt angle $\varphi$ can be seen in Fig.~\ref{RelativePhaseDependence}. For angles $\varphi = -22^\circ$, $-1^\circ$ and $19^\circ$, the relative phase between components is equal to $\phi' = -2\pi$, $0$ and $2\pi$ while for $\varphi = -15^\circ$ and $12^\circ$, the relative phase between components is equal to $\phi' = -\pi$ and $\pi$.

\section{Entanglement characterization, QST and Bell Test}
\label{sec:exp}

Once we have the setup optimally aligned, we can characterize and perform a Bell test in all four Bell states. As explained above, the evaluation of the 
CHSH inequality requires measuring several correlation 
functions, Eq.~(\ref{E}), which contain coincidence 
probabilities among the detectors. Thus, to start 
characterizing the correlations arising in our 
detectors we compare quantum mechanical preditions 
to our data for a couple such correlations. 

In particular, we generate the plots 
$C(0^\circ,\theta)$ and $C(45^\circ,\theta)$ to observe 
if the experimental results follows the results 
predicted by quantum mechanics. The $C(0^\circ,\theta)$ 
and $C(45^\circ,\theta)$ plots shown in Fig.~\ref{Phi_Figures} 
and \ref{Psi_Figures} are obtained measuring the coincidence 
counts of pairs of photons in the state $\ket{V_{0^\circ}} \otimes \ket{V_{\theta}}$ or $\ket{V_{45^\circ}} \otimes \ket{V_{\theta}}$ respectively, while varying the angle $\theta$ at which we measure the state of the second photon from $0^\circ$ to $180^\circ$ in steps of $10^\circ$.

Then, we also conduct quantum state tomography for each case 
and determine the fidelity between our state and the 
desired state, see Table~\ref{BellTest_Fidelity}. The results 
are shown in Fig.~\ref{Phi_Figures}  and~\ref{Psi_Figures}, 
where the Real Part and Imaginary Part in the diagrams refer 
to the weights of the real and imaginary parts, respectively, 
in the density matrix associated with the state of our photons. 
The coefficient associated with each of these weights can 
be determined by observing the labels on the axes of the diagrams.


Finally, to prove that we have indeed an entangled state, 
we perform the Bell test. For each of the states, we 
obtain a 
violation of the Bell inequalities with at least 40 
standard deviations from the maximum classical value of 
$| \langle S \rangle | = 2$, see Table \ref{BellTest_Fidelity}. 
As each measurement takes 30 seconds, the 
time needed for conducting a Bell 
test using Setup 1 is 20~minutes and 5 minutes with Setup 2. The resulting value for the state $\ket{\Phi^{+}}$ is $\langle S \rangle = 2.730 \pm 0.015$.

For the tomography and the calculation of the fidelity, 
we use only the number of coincidence counts, while 
for obtaining the value of the Bell inequality, we use 
the difference between coincidence counts and 
accidental counts, as described in Sec.~\ref{sec:app3}.

\begin{table}[H]
\begin{ruledtabular}
\begin{tabular}{cccc}
\textrm{State}&
$\langle S \rangle$ & $\langle S' \rangle$ &
\multicolumn{1}{c}{Fidelity}\\
\colrule
 $\ket{\Phi^{+}}$ & $2.765 \pm 0.018$ & $0.022 \pm 0.018$ & $0.945 \pm 0.005$ \\ 
\hline

$\ket{\Phi^{-}}$ & $0.101 \pm 0.016$ & $2.745 \pm 0.016$ & $0.918 \pm 0.005$ \\ 
\hline

$\ket{\Psi^{+}}$ & $-0.055 \pm 0.019$ & $-2.806 \pm 0.019$ & $0.881 \pm 0.005$ \\ 
\hline

$\ket{\Psi^{-}}$ & $-2.804 \pm 0.018$ & $-0.053 \pm 0.018$ & $0.954 \pm 0.005$\\
\end{tabular}
\end{ruledtabular}
\caption{Bell test and fidelity results for each of states.}
\label{BellTest_Fidelity}
\end{table}

\begin{widetext}

\begin{figure}[p]
\begin{minipage}{.45\textwidth}
    \centering
    \includegraphics[width=1\linewidth]{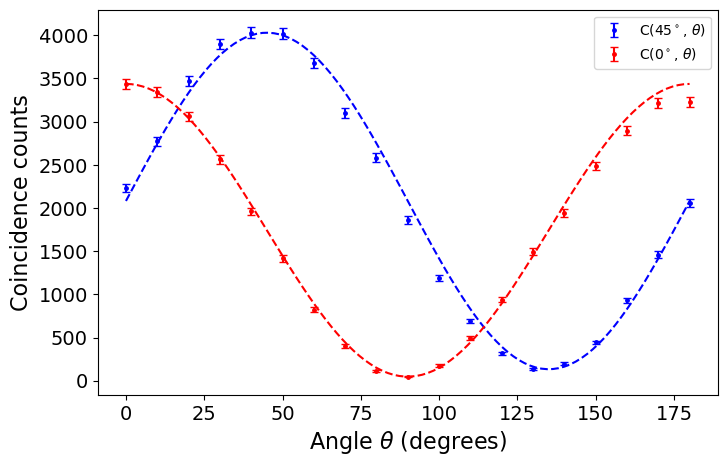}
    \includegraphics[width=1\linewidth]{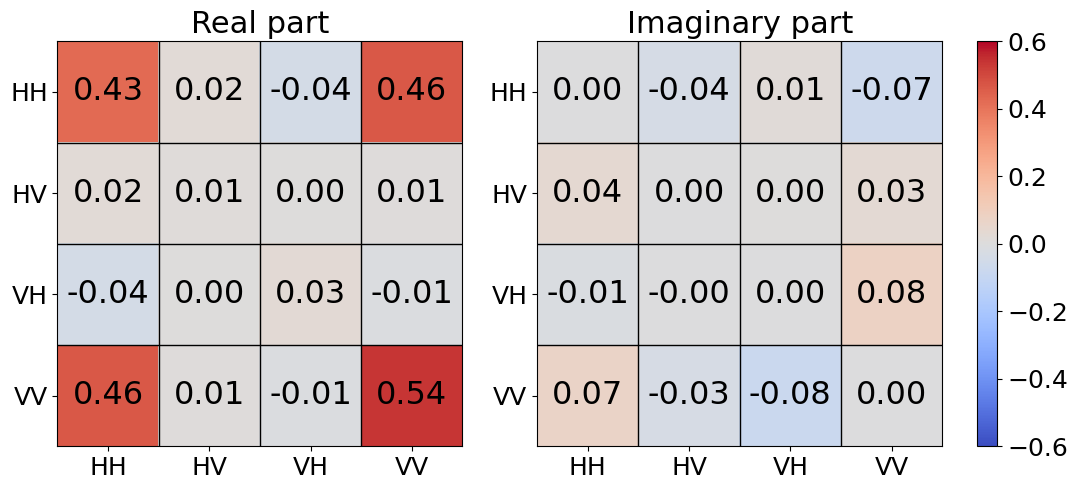}
\end{minipage}
\hspace{0.5cm}
\begin{minipage}{.45\textwidth}
  \centering
  \includegraphics[width=1\linewidth]{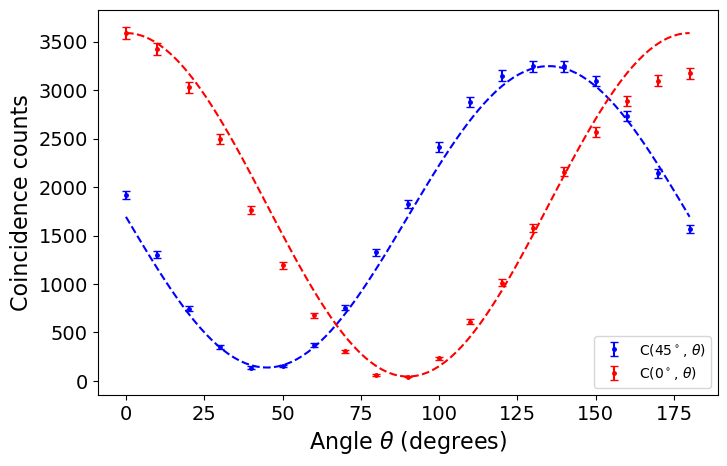}
  \includegraphics[width=1\linewidth]{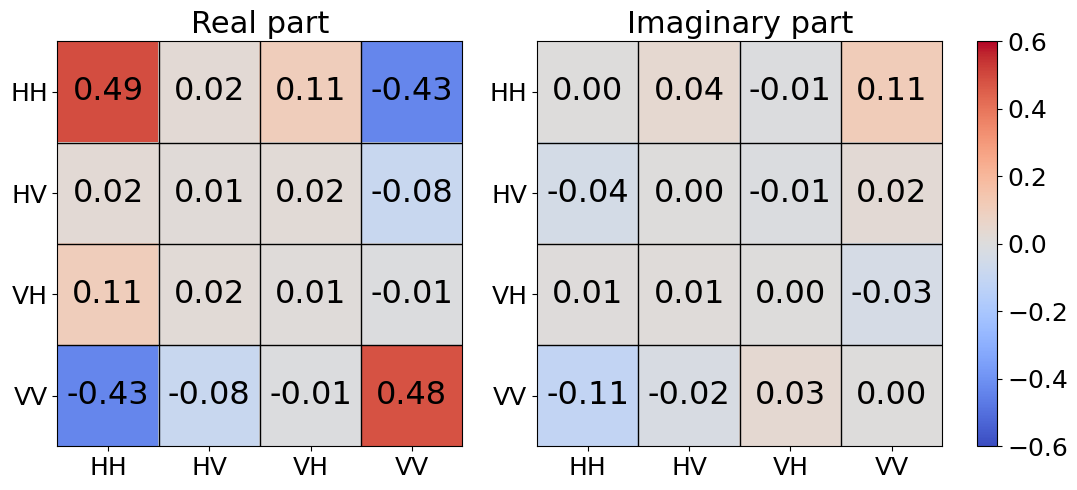}
\end{minipage}
\caption{ $C(0^\circ,\theta)$, $C(45^\circ,\theta)$ graphs and tomography for the states $\ket{\Phi^{+}}$ (left) and $\ket{\Phi^{-}}$ (right). The dashed lines correspond to the 
quantum mechanics predictions, 
$C(\theta_1,\theta_2) \propto \cos^2(\theta_1-\theta_2)$ and 
$C(\theta_1,\theta_2) \propto \cos^2(\theta_1+\theta_2)$, 
for $\ket{\Phi^{+}}$ and $\ket{\Phi^{-}}$, respectively.  See Sec.~\ref{sec:app2}.}
\label{Phi_Figures}
\end{figure}

\begin{figure}[p]
\begin{minipage}{.45\textwidth}
  \centering
   \includegraphics[width=1\linewidth]{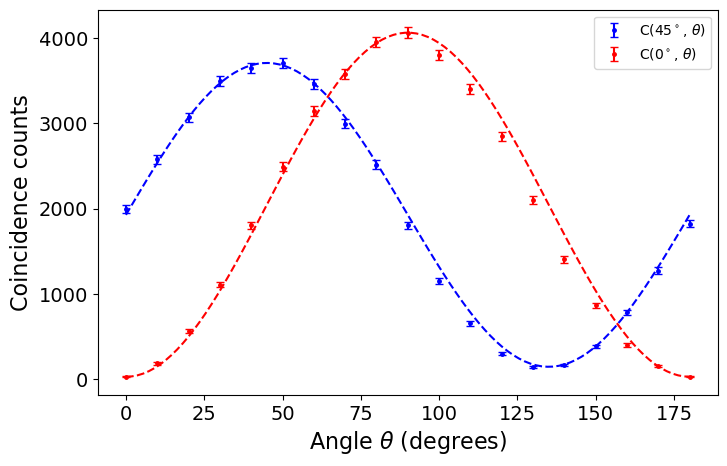}
  \includegraphics[width=1\linewidth]{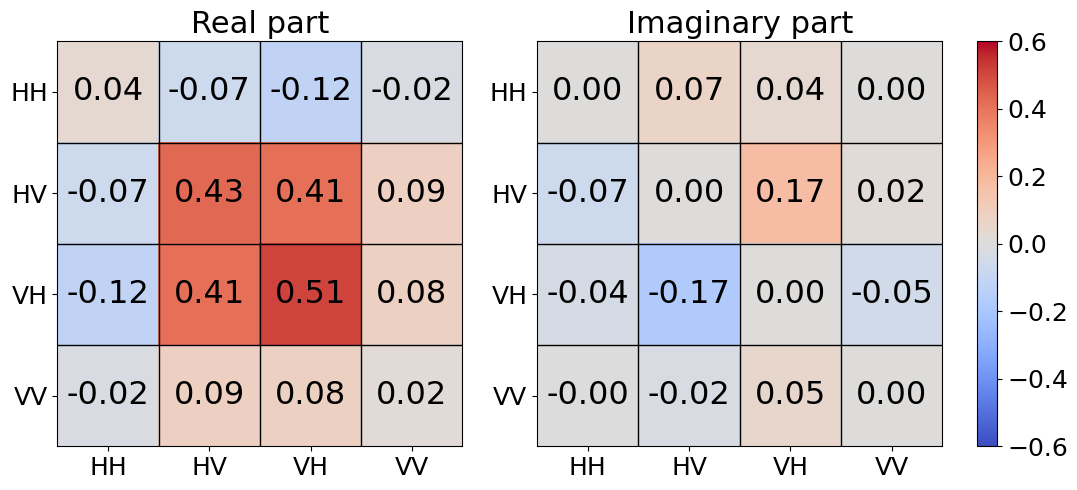}
\end{minipage}
\hspace{0.5cm}
\begin{minipage}{.45\textwidth}
  \centering
  \includegraphics[width=1\linewidth]{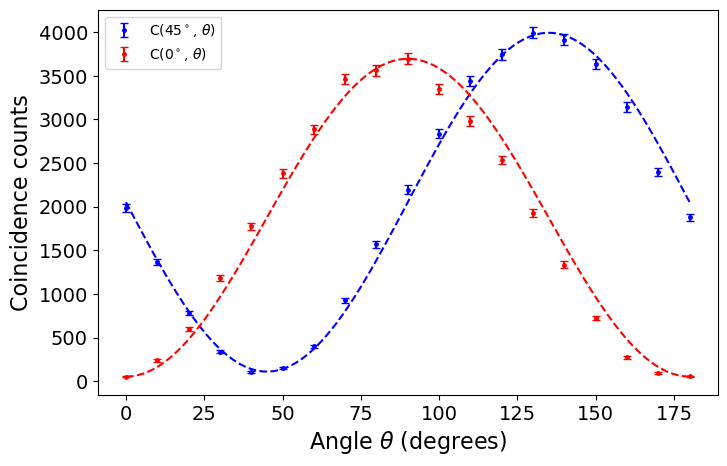}
  \includegraphics[width=1\linewidth]{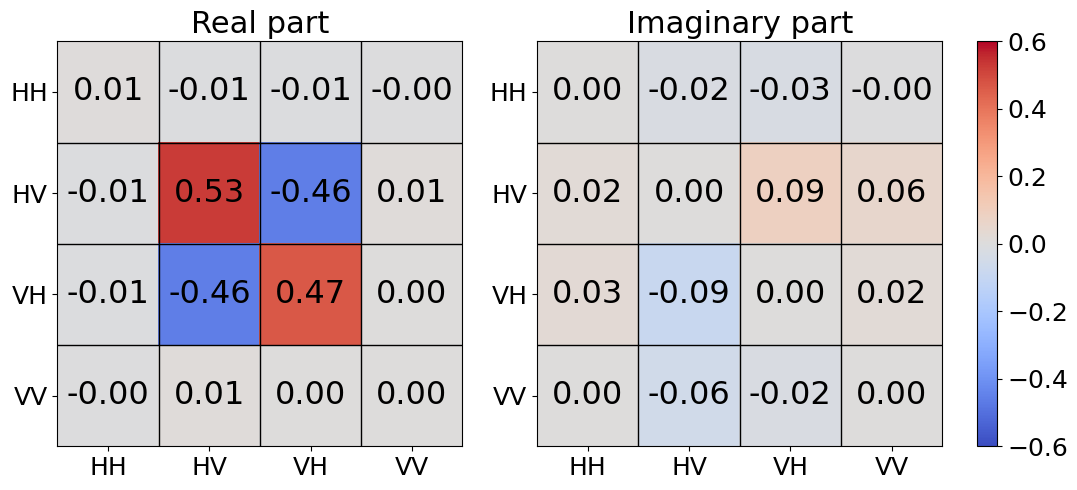}
\end{minipage}
\caption{ $C(0^\circ,\theta)$, $C(45^\circ,\theta)$ graphs and tomography for the states $\ket{\Psi^{+}}$ (left) and $\ket{\Psi^{-}}$ (right). The dashed lines correspond to the 
quantum mechanics predictions, 
$C(\theta_1,\theta_2) \propto \sin^2(\theta_1+\theta_2)$ and 
$C(\theta_1,\theta_2) \propto \sin^2(\theta_1-\theta_2)$, 
for $\ket{\Psi^{+}}$ and $\ket{\Psi^{-}}$, respectively. See Sec.~\ref{sec:app2}.}
\label{Psi_Figures}
\end{figure}

\end{widetext}
\clearpage

\section{Conclusions}
\label{sec:conclussion}

We have presented a new experimental laboratory aimed at 
the undergraduate and master level to study quantum entangled 
photons. Two different setups have been described, 
which differ on the photon detection part. The photon production 
is in both cases through two BBO crystals. The photon detection 
is either performed with two single photon detectors or with 
four, allowing in the latter case to reduce the measurement 
time by a factor four. 

The photons are collected by means of optical fibers mounted 
on custom made rails, thus ensuring an easy and robust 
alignment. The procedure to assemble and align the system 
from scratch, which has proven key in our experience, has been presented, thus providing a direct guide
to future undergrad students in quantum science and technology 
laboratories worldwide. 

The experiments which can be conducted are manifold. First, 
one can produce any of the well know Bell states, in our case 
produced with a fidelity higher that 88\%. The full tomography 
of the states can be performed, thus confirming that the 
desired two-photon quantum state has been produced. Besides, one 
can also perform correlated measurements within the two 
photons, which can be directly confronted with quantum mechanical predictions. 

Bell tests tailored for the different Bell states can also 
be performed. In our case, we measured violations of the 
corresponding inequalities by more than 40 standard deviations. 
With this setup, after alignment, a Bell test can be conducted 
in less than an hour.  

All of this, combined with the fact that the total 
cost of components required to assemble both setups is 
approximately twenty thousand euros (see Sec.~\ref{sec:app1}), 
makes this experiment accessible to laboratories with limited resources. It also brings these 
type of demonstrations closer to undergraduate students, high 
school students, or even a broader audience. This facilitates the dissemination of key concepts in quantum mechanics beyond universities and specialized research groups.

\begin{acknowledgements}
We thank Marti Pedemonte and Alejandro Jaramillo for their  contribution to early versions of the current setup, Morgan Mitchell for his support and suggestions over the years. We acknowledge useful and constructive discussions with Radek Lapkiewicz, Lluís 
Garrido and Hector Briongos. This study was supported by MCIN with funding from European 
Union NextGenerationEU(PRTR-C17.I1) and by Generalitat de Catalunya. 
We acknowledge funding from Grant No.~PID2020-114626GB-I00 
by MCIN/AEI/10.13039/5011 00011033 and "Unit of Excellence 
Mar\'ia de Maeztu 2020-2023” award to the Institute of Cosmos 
Sciences, Grant CEX2019-000918-M by MCIN/AEI/10.13039/501100011033, 
and Grants 2021SGR01095, 2021SGR00242 
and 2021SGR01108 by Generalitat de Catalunya, and Grant 101002460-DEEP by European Research Council.  This project has received funding from the 
European Union’s Digital Europe Programme under grant agreement no.
101084035.
\end{acknowledgements}
\bibliographystyle{apsrev}
\bibliography{apssamp}

\begin{widetext}

\appendix

\clearpage

\section{List of material}
\label{sec:app1}

For any reader interested in replicating either of the two setups, here is an inventory of useful information for acquiring all the necessary elements. The total cost of all these elements amounts to around 20.000 euros.

\begin{table} [h]
\begin{tabular}{|l|l|l|c|}
\colrule
Description of the product & Reference and Company & Price (eur.) & Quantity \\
\colrule
Precision kinematic mount & \textsf{KS1, ThorLabs} & 90.51 & $4+2_{2}$ \\
Fiber-coupling lenses &\textsf{F810FC-780, ThorLabs} & 267.73 & $2+2_{2}$ \\
Goniometer &\textsf{RP01/M, ThorLabs} & 101.24 & $1$ \\
Rotation Mount &\textsf{RSP1X225/M, ThorLabs} & 141.46 & $4$ \\
Rotation Mount 30mm cage system &\textsf{CRM1T/M, ThorLabs}  & 87.20 & $1$ \\
BBO crystals mount &\textsf{KS1RS, ThorLabs} & 250.90 & $1$ \\
PBS mount &\textsf{KM200PM/M, ThorLabs} & 127.41 & $2_{2}$ \\

4-channel detector &\textsf{SPCM-AQ4C, Excelitas Technologies} & 13456.21 & $1$ \\
Current and Temperature Controllers for Laser Diodes &\textsf{LTC56A/M, ThorLabs} & 2847.77 & $1$ \\

Laser diode 404nm 400mW &\textsf{L404P400M, ThorLabs} & 684.60 & $1$ \\
Mirrors &\textsf{BB1-E02, ThorLabs} & 73.83 & $2$ \\
BBO crystals &\textsf{EKSMA Optics} & 1540.00 & $1$ \\ 
Bandpass Filter 405nm &\textsf{FBH405-10, ThorLabs} & 149.69 & $1$ \\
Bandpass Filter 800nm &\textsf{FBH800-40, ThorLabs} & 149.69 & $2 + 2_{2}$ \\
“Infrared” Polarizers &\textsf{LPNIRE100-B, ThorLabs} & 116.46 & $2_{1}$ \\
Quarter-wave Plate 808nm &\textsf{WPQ05M-808, ThorLabs} & 461.09 & $2$ \\
Half-wave Plate 808nm &\textsf{WPH05M-808, ThorLabs} & 461.09 & $2_{2}$ \\
Half-wave Plate 405nm &\textsf{WPH05M-405, ThorLabs} & 461.09 & $2$ \\
Polarizing Beamsplitter (PBS) &\textsf{PBS252, ThorLabs} & 235.40 & $2_{2}$ \\
Development board for the coincidence circuit &\textsf{NUCLEO-F756ZG, STMicroelectronics} & 22.82 & $1$ \\
\colrule

\end{tabular}
\caption{Detailed equipment used in the setups. The subscripts 1 and 2 denote the quantities of elements required solely for the construction of setups 1 and 2, respectively. \label{tab:eq}}
\end{table}

\section{Stokes coefficients}
\label{sec:app-stokes}

The general equation for each of the Stokes coefficients of Eq.~(\ref{Tomo}) is

\begin{equation}
    S_{ij} = \rm{Tr}(\sigma_i \otimes \sigma_j \cdot \rho),
\end{equation}

\noindent where $\rho$ is the state of our pair of photons. Thus, the explicit expression for the Stokes 
coefficients is, 
\begin{align*}
    &S_{00} = P_{\ket{HH}} + P_{\ket{HV}} + P_{\ket{VH}} + P_{\ket{VV}},
    \\
    &S_{01} = P_{\ket{HD}} - P_{\ket{HA}} + P_{\ket{VD}} - P_{\ket{VA}},
    \\
    &S_{02} = P_{\ket{HL}} - P_{\ket{HR}} + P_{\ket{VL}} - P_{\ket{VR}},
    \\
    &S_{03} = P_{\ket{HH}} - P_{\ket{HV}} + P_{\ket{VH}} - P_{\ket{VV}},
    \\
    &S_{10} = P_{\ket{DH}} + P_{\ket{DV}} - P_{\ket{AH}} - P_{\ket{AV}},
    \\
    &S_{11} = P_{\ket{DD}} - P_{\ket{DA}} - P_{\ket{AD}} + P_{\ket{AA}},
    \\
    &S_{12} = P_{\ket{DL}} - P_{\ket{DR}} - P_{\ket{AL}} + P_{\ket{AR}},
    \\
    &S_{13} = P_{\ket{DH}} - P_{\ket{DV}} - P_{\ket{AH}} + P_{\ket{AV}},
    \\
    &S_{20} = P_{\ket{LH}} + P_{\ket{LV}} - P_{\ket{RH}} - P_{\ket{RV}},
    \\
    &S_{21} = P_{\ket{LD}} - P_{\ket{LA}} - P_{\ket{RD}} + P_{\ket{RA}},
    \\
    &S_{22} = P_{\ket{LL}} - P_{\ket{LR}} - P_{\ket{RL}} + P_{\ket{RR}},
    \\
    &S_{23} = P_{\ket{LH}} - P_{\ket{LV}} - P_{\ket{RH}} + P_{\ket{RV}},
    \\
    &S_{30} = P_{\ket{HH}} + P_{\ket{HV}} - P_{\ket{VH}} - P_{\ket{VV}},
    \\
    &S_{31} = P_{\ket{HD}} - P_{\ket{HA}} - P_{\ket{VD}} + P_{\ket{VA}},
    \\
    &S_{32} = P_{\ket{HL}} - P_{\ket{HR}} - P_{\ket{VL}} + P_{\ket{VR}},
    \\
    &S_{33} = P_{\ket{HH}} - P_{\ket{HV}} - P_{\ket{VH}} + P_{\ket{VV}}\,.
\end{align*}

\section{Computation of of $P_{\ket{V_{\theta_1}V_{\theta_2}}}$ for all four Bell States}
\label{sec:app2}

We define $P_{\ket{V_{\theta_1}V_{\theta_2}}}$ as the probability of finding the two photons in the state $\ket{V_{\theta_1}}\otimes\ket{V_{\theta_2}}$. The values of this probabilities for each one of the states are:

\begin{itemize}
    \item For $\ket{\Phi^{+}} = \frac{1}{\sqrt{2}} (\ket{HH} + \ket{VV})$:

\begin{align}
        &P_{\ket{V_{\theta_1}V_{\theta_2}}} = |\langle \Phi^{+} | V_{\theta_1}V_{\theta_2} \rangle|^2 \nonumber\\
        &= |\frac{1}{\sqrt{2}} (\bra{HH} + \bra{VV}) \cdot (\sin\theta_1 \sin\theta_2 \ket{HH} + \cos\theta_1 \cos\theta_2 \ket{VV} -\sin\theta_1 \cos\theta_2 \ket{HV} -\cos\theta_1 \sin\theta_2 \ket{VH})|^2 \nonumber  \\
        &= |\frac{1}{\sqrt{2}} (\sin\theta_1 \sin\theta_2 + \cos\theta_1 \cos\theta_2)|^2 = |\frac{1}{\sqrt{2}} \cos(\theta_1 - \theta_2)|^2 = \frac{1}{2} \cos^2(\theta_1 - \theta_2)
    \end{align}

    \item For $\ket{\Phi^{-}} = \frac{1}{\sqrt{2}} (\ket{HH} - \ket{VV})$:
\begin{align}
        &P_{\ket{V_{\theta_1}V_{\theta_2}}} = |\langle \Phi^{-} | V_{\theta_1}V_{\theta_2} \rangle|^2  \nonumber\\
        &= |\frac{1}{\sqrt{2}} (\bra{HH} - \bra{VV}) \cdot (\sin\theta_1 \sin\theta_2 \ket{HH} + \cos\theta_1 \cos\theta_2 \ket{VV} -\sin\theta_1 \cos\theta_2 \ket{HV} -\cos\theta_1 \sin\theta_2 \ket{VH})|^2 \nonumber\\
        &= |\frac{1}{\sqrt{2}} (\sin\theta_1 \sin\theta_2 - \cos\theta_1 \cos\theta_2)|^2 = |\frac{1}{\sqrt{2}} \cos(\theta_1 + \theta_2)|^2 = \frac{1}{2} \cos^2(\theta_1 + \theta_2)
    \end{align}

    \item For $\ket{\Psi^{+}} = \frac{1}{\sqrt{2}} (\ket{HV} + \ket{VH})$:

\begin{align}
        &P_{\ket{V_{\theta_1}V_{\theta_2}}} = |\langle \Psi^{+} | V_{\theta_1}V_{\theta_2} \rangle|^2 \nonumber\\
        &= |\frac{1}{\sqrt{2}} (\bra{HV} + \bra{VH}) \cdot (\sin\theta_1 \sin\theta_2 \ket{HH} + \cos\theta_1 \cos\theta_2 \ket{VV} -\sin\theta_1 \cos\theta_2 \ket{HV} -\cos\theta_1 \sin\theta_2 \ket{VH})|^2 \nonumber \\
        &= |\frac{1}{\sqrt{2}} (\sin\theta_1 \cos\theta_2 + \cos\theta_1 \sin\theta_2)|^2 = |\frac{1}{\sqrt{2}} \sin(\theta_1 + \theta_2)|^2 = \frac{1}{2} \sin^2(\theta_1 + \theta_2)
    \end{align}

    \item For $\ket{\Psi^{-}} = \frac{1}{\sqrt{2}} (\ket{HV} - \ket{VH})$:

 \begin{align}
        &P_{\ket{V_{\theta_1}V_{\theta_2}}} = |\langle \Psi^{-} | V_{\theta_1}V_{\theta_2} \rangle|^2  \nonumber\\
        &= |\frac{1}{\sqrt{2}} (\bra{HV} - \bra{VH}) \cdot (\sin\theta_1 \sin\theta_2 \ket{HH} + \cos\theta_1 \cos\theta_2 \ket{VV} -\sin\theta_1 \cos\theta_2 \ket{HV} -\cos\theta_1 \sin\theta_2 \ket{VH})|^2 \nonumber \\
        &= |\frac{1}{\sqrt{2}} (\sin\theta_1 \cos\theta_2 - \cos\theta_1 \sin\theta_2)|^2 = |\frac{1}{\sqrt{2}} \sin(\theta_1 - \theta_2)|^2 = \frac{1}{2} \sin^2(\theta_1 - \theta_2)
    \end{align}
    
\end{itemize}

We can summarize all this in Table~\ref{tab1}. 
\begin{table}[b]
\centering
\begin{tabular}{|c|c|c|}
\hline
Bell State & $P_{\ket{V_{\theta_1}V_{\theta_2}}}$ \\ \hline
$\ket{\Phi^{+}}$ & $\frac{1}{2} \cos^2(\theta_1 - \theta_2)$  \\ \hline
$\ket{\Phi^{-}}$ & $\frac{1}{2} \cos^2(\theta_1 + \theta_2)$  \\ \hline
$\ket{\Psi^{+}}$ & $\frac{1}{2} \sin^2(\theta_1 + \theta_2)$  \\ \hline
$\ket{\Psi^{-}}$ & $\frac{1}{2} \sin^2(\theta_1 - \theta_2)$  \\ \hline
\end{tabular}
\caption{Values of $P_{\ket{V_{\theta_1}V_{\theta_2}}}$ for each one of the Bell states.
\label{tab1}}
\end{table}

\section{Data for each Bell Test}
\label{sec:app3}

In this section, we present all the data obtained for each one of the Bell test performed to the different Bell states using the setup depicted in Fig.~\ref{Setup1}. We also provide the data for the Bell test achieved with the photons in the state $\ket{\Phi^{+}}$, using the setup presented in Fig.~\ref{Setup2}. In all theese graphs, the angles $\alpha$ and $\beta$ stands for the angles at which Alice ($\ket{V_\alpha}$)  and Bob ($\ket{V_\beta}$) measure their respective photons. Each of these measurements was taken using a time interval of 30 seconds.

\begin{table}[H]
\centering
\begin{tabular}{l@{\hskip 15pt} l@{\hskip 15pt} l@{\hskip 15pt} l@{\hskip 15pt} l@{\hskip 15pt} l}
\hline\hline
\textbf{$\alpha$ ($^\circ$)} & \textbf{$\beta$ ($^\circ$)} & \textbf{$N_a$} & \textbf{$N_b$} & \textbf{$N_c$} & \textbf{$N_{acc}$} \\ \hline
45 & 22.5 & 242324 & 126944 & 3025 & 92.28 \\ \hline
45 & -22.5 & 241250 & 125920 & 377 & 91.13 \\ \hline
45 & -67.5 & 244869 & 142160 & 1242 & 104.43 \\ \hline
45 & -112.5 & 250568 & 145332 & 3959 & 109.25 \\ \hline
0 & 22.5 & 231257 & 137000 & 3432 & 95.05 \\ \hline
0 & -22.5 & 235528 & 132632 & 2975 & 93.72 \\ \hline
0 & -67.5 & 226909 & 138908 & 310 & 94.56 \\ \hline
0 & -112.5 & 226724 & 143352 & 1028 & 97.50 \\ \hline
-45 & 22.5 & 234822 & 135900 & 763 & 95.74 \\ \hline
-45 & -22.5 & 234676 & 132592 & 3684 & 93.35 \\ \hline
-45 & -67.5 & 233508 & 145184 & 3376 & 101.70 \\ \hline
-45 & -112.5 & 229828 & 150008 & 545 & 103.43 \\ \hline
-90 & 22.5 & 224928 & 129820 & 394 & 87.60 \\ \hline
-90 & -22.5 & 222538 & 123984 & 927 & 82.77 \\ \hline
-90 & -67.5 & 217928 & 135868 & 3728 & 88.83 \\ \hline
-90 & -112.5 & 223545 & 141888 & 3031 & 95.16 \\ \hline\hline
\end{tabular}
\caption{Data obtained for the computation of the Bell inequality using pairs of photons in the $\ket{\Phi^{+}}$.}
\end{table}

\begin{table}[H]
\centering
\begin{tabular}{l@{\hskip 15pt} l@{\hskip 15pt} l@{\hskip 15pt} l@{\hskip 15pt} l@{\hskip 15pt} l}
\hline\hline
\textbf{$\alpha$ ($^\circ$)} & \textbf{$\beta$ ($^\circ$)} & \textbf{$N_a$} & \textbf{$N_b$} & \textbf{$N_c$} & \textbf{$N_{acc}$} \\ \hline
45 & 22.5 & 220634 & 157440 & 721 & 104.21 \\ \hline
45 & -22.5 & 226207 & 151083 & 3995 & 102.53 \\ \hline
45 & -67.5 & 226708 & 173348 & 4562 & 117.90 \\ \hline
45 & -112.5 & 221408 & 182304 & 1134 & 121.09 \\ \hline
0 & 22.5 & 208526 & 157300 & 4001 & 98.40 \\ \hline
0 & -22.5 & 214617 & 149412 & 3915 & 96.20 \\ \hline
0 & -67.5 & 208804 & 169832 & 565 & 106.38 \\ \hline
0 & -112.5 & 213806 & 182984 & 1001 & 117.37 \\ \hline
-45 & 22.5 & 245625 & 157568 & 4293 & 116.11 \\ \hline
-45 & -22.5 & 246193 & 149052 & 683 & 110.09 \\ \hline
-45 & -67.5 & 240321 & 168744 & 1417 & 121.66 \\ \hline
-45 & -112.5 & 241023 & 178476 & 4720 & 129.05 \\ \hline
-90 & 22.5 & 259133 & 157176 & 852 & 122.19 \\ \hline
-90 & -22.5 & 259584 & 149428 & 835 & 116.37 \\ \hline
-90 & -67.5 & 254570 & 174292 & 4523 & 133.11 \\ \hline
-90 & -112.5 & 259205 & 182424 & 4554 & 141.86 \\ \hline
\end{tabular}
\caption{Data obtained for the computation of the Bell inequality using pairs of photons in the $\ket{\Phi^{-}}$ state.}
\end{table}

\begin{table}[H]
\centering
\begin{tabular}{l@{\hskip 15pt} l@{\hskip 15pt} l@{\hskip 15pt} l@{\hskip 15pt} l@{\hskip 15pt} l}
\hline\hline
\textbf{$\alpha$ ($^\circ$)} & \textbf{$\beta$ ($^\circ$)} & \textbf{$N_a$} & \textbf{$N_b$} & \textbf{$N_c$} & \textbf{$N_{acc}$} \\ \hline
45 & 22.5 & 202953 & 140256 & 2997 & 85.40 \\ \hline
45 & -22.5 & 209729 & 140292 & 589 & 88.27 \\ \hline
45 & -67.5 & 199521 & 132980 & 854 & 79.60 \\ \hline
45 & -112.5 & 199216 & 138172 & 3409 & 82.58 \\ \hline
0 & 22.5 & 195788 & 135044 & 520 & 79.32 \\ \hline
0 & -22.5 & 191147 & 126572 & 716 & 72.58 \\ \hline
0 & -67.5 & 196727 & 130908 & 3435 & 77.26 \\ \hline
0 & -112.5 & 191245 & 138028 & 3167 & 79.19 \\ \hline
-45 & 22.5 & 185983 & 135516 & 648 & 75.61 \\ \hline
-45 & -22.5 & 187608 & 130676 & 3189 & 73.55 \\ \hline
-45 & -67.5 & 188891 & 133528 & 2898 & 75.67 \\ \hline
-45 & -112.5 & 185914 & 136744 & 470 & 76.27 \\ \hline
-90 & 22.5 & 179726 & 137788 & 2973 & 74.29 \\ \hline
-90 & -22.5 & 178394 & 131576 & 2709 & 70.42 \\ \hline
-90 & -67.5 & 179378 & 131052 & 401 & 70.52 \\ \hline
-90 & -112.5 & 175836 & 133828 & 642 & 70.60 \\ \hline
\end{tabular}
\caption{Data obtained for the computation of the Bell inequality using pairs of photons in the $\ket{\Psi^{+}}$ state.}
\end{table}

\begin{table}[H]
\centering
\begin{tabular}{l@{\hskip 15pt} l@{\hskip 15pt} l@{\hskip 15pt} l@{\hskip 15pt} l@{\hskip 15pt} l}
\hline\hline
\textbf{$\alpha$ ($^\circ$)} & \textbf{$\beta$ ($^\circ$)} & \textbf{$N_a$} & \textbf{$N_b$} & \textbf{$N_c$} & \textbf{$N_{acc}$} \\ \hline
45 & 22.5 & 205843 & 152544 & 1026 & 94.20 \\ \hline
45 & -22.5 & 202783 & 146180 & 3625 & 88.93 \\ \hline
45 & -67.5 & 203926 & 136128 & 2977 & 83.28 \\ \hline
45 & -112.5 & 203209 & 140200 & 370 & 85.47 \\ \hline
0 & 22.5 & 199956 & 155508 & 476 & 93.28 \\ \hline
0 & -22.5 & 198214 & 149576 & 728 & 88.94 \\ \hline
0 & -67.5 & 194361 & 139668 & 3686 & 81.44 \\ \hline
0 & -112.5 & 195118 & 143816 & 3027 & 84.18 \\ \hline
-45 & 22.5 & 209593 & 155160 & 3400 & 97.56 \\ \hline
-45 & -22.5 & 209790 & 149908 & 643 & 94.35 \\ \hline
-45 & -67.5 & 207594 & 136140 & 740 & 84.79 \\ \hline
-45 & -112.5 & 202074 & 142580 & 3403 & 86.44 \\ \hline
-90 & 22.5 & 197175 & 147948 & 3358 & 87.51 \\ \hline
-90 & -22.5 & 202596 & 146072 & 3212 & 88.78 \\ \hline
-90 & -67.5 & 204167 & 135736 & 438 & 83.14 \\ \hline
-90 & -112.5 & 202654 & 143952 & 860 & 87.52 \\ \hline
\end{tabular}
\caption{Data obtained for the computation of the Bell inequality using pairs of photons in the $\ket{\Psi^{-}}$ state.}
\end{table}

\begin{table}[H]
\centering
\begin{tabular}{l@{\hskip 15pt} l@{\hskip 15pt} l@{\hskip 15pt} l@{\hskip 15pt} l@{\hskip 15pt} l}
\hline\hline
\textbf{$\alpha$ ($^\circ$)} & \textbf{$\beta$ ($^\circ$)} & \textbf{$N_a$} & \textbf{$N_b$} & \textbf{$N_c$} & \textbf{$N_{acc}$} \\ \hline
45 & 22.5 & 305415 & 248380 & 5302 & 227.57 \\ \hline
45 & -22.5 & 280234 & 237452 & 1709 & 199.62 \\ \hline
45 & -67.5 & 190449 & 247220 & 859 & 141.24 \\ \hline
45 & -112.5 & 210298 & 247104 & 5626 & 155.89 \\ \hline
0 & 22.5 & 305380 & 320864 & 6682 & 293.95 \\ \hline
0 & -22.5 & 288215 & 312288 & 6478 & 270.01 \\ \hline
0 & -67.5 & 194873 & 321212 & 1267 & 187.78 \\ \hline
0 & -112.5 & 201912 & 301124 & 1538 & 182.40 \\ \hline
-45 & 22.5 & 294989 & 231004 & 2013 & 204.43 \\ \hline
-45 & -22.5 & 280329 & 230040 & 5343 & 193.46 \\ \hline
-45 & -67.5 & 184691 & 232300 & 4809 & 128.71 \\ \hline
-45 & -112.5 & 202687 & 237160 & 604 & 144.20 \\ \hline
-90 & 22.5 & 292146 & 164976 & 729 & 144.59 \\ \hline
-90 & -22.5 & 278696 & 170320 & 513 & 142.40 \\ \hline
-90 & -67.5 & 174801 & 158404 & 4295 & 83.06 \\ \hline
-90 & -112.5 & 201492 & 164912 & 4770 & 99.68 \\ \hline
\end{tabular}
\caption{Data obtained for the computation of the Bell inequality with the setup depicted in Fig.~\ref{Setup2}, using pairs of photons in the $\ket{\Phi^{+}}$ state.}
\end{table}

Assuming that the counts follow a Poisson distribution, the error corresponding to the number of counts detected in a certain time interval is equal to the square root of that value. That is:

\begin{equation}
    \sigma_{N_a} = \sqrt{N_a}\,,
\end{equation}
\begin{equation}
    \sigma_{N_b} = \sqrt{N_b}\,,
\end{equation}
\begin{equation}
    \sigma_{N_c} = \sqrt{N_c}\,.
\end{equation}

\noindent Therefore, performing error propagation in Eq.~(\ref{N_acc}), we have that the error in $N_{acc}$ is,

\begin{equation}
    \sigma_{N_{acc}} = \frac{\tau}{T}\cdot \sqrt{N_b^2 \cdot \sigma_{N_a}^2 + N_a^2 \cdot \sigma_{N_b}^2} = \frac{\tau}{T}\cdot \sqrt{N_b^2 \cdot N_a + N_a^2 \cdot N_b},
\end{equation}

\noindent where we have assumed that the values $T$ and $\tau$ have no associated error. For simplicity, we define $N_{VV} = N_{\ket{V_\alpha V_\beta}} = N_c(\ket{V_\alpha V_\beta}) - N_{acc}(\ket{V_\alpha V_\beta})$ as the number of detected coincidence counts minus the number of accidental coincidence counts and $\sigma_{N_{VV}} = \sigma_{N}(\ket{V_\alpha V_\beta})$ to their associated error,
\begin{equation}
    \sigma_{N_{VV}} = \sqrt{\sigma^2_{N_c}(\ket{V_\alpha V_\beta}) + \sigma^2_{N_{acc}}(\ket{V_\alpha V_\beta})}.
\end{equation}

\noindent Then, the equation for the different probabilities is 
\begin{equation}
\begin{split}
    P_{\ket{V_\alpha V_\beta}} &= \frac{N_{\ket{V_\alpha V_\beta}}}{N_{\ket{V_\alpha V_\beta}} + N_{\ket{V_{\alpha-90^\circ} V_\beta}} + N_{\ket{V_\alpha V_{\beta-90^\circ}}} + N_{\ket{V_{\alpha-90^\circ} V_{\beta-90^\circ}}}} \\
    &= \frac{N_{\ket{V_\alpha V_\beta}}}{N_{\ket{V_\alpha V_\beta}} + N_{\ket{H_{\alpha} V_\beta}} + N_{\ket{V_\alpha H_{\beta}}} + N_{\ket{H_{\alpha} H_{\beta}}}} \\
    &= \frac{N_{VV}}{N_{VV} + N_{HV} + N_{VH} + N_{HH}},
\end{split}
\end{equation}
\noindent where, from the Eq.~(\ref{VaHa}), we extract that $\ket{V_{\gamma-90^\circ}} = \ket{H_{\gamma}}$. The error associated to the probability $P_{\ket{V_\alpha V_\beta}}$ can be written as 
\begin{equation}
\begin{split}
    \sigma_{P_{\ket{V_\alpha V_\beta}}} &=  \frac{1}{\left( N_{VV} + N_{HH} + N_{VH} + N_{HH} \right)^2} \cdot \sqrt{ N^2_{VV} \cdot (\sigma_{N_{VH}}^2 + \sigma_{N_{HV}}^2 + \sigma_{N_{HH}}^2 ) + (N_{HV} + N_{VH} + N_{HH})^2 \cdot \sigma_{N_{VV}}^2 }.
\end{split}
\end{equation}

\noindent  For the correlation functions, defined as in Eq.~(\ref{E}), we have that the error is 
\begin{equation}
\begin{split}
    \sigma_{E(\alpha, \beta)}  = \sqrt{\sigma^2_{P_{\ket{V_\alpha V_\beta}}} + \sigma^2_{P_{\ket{V_\alpha H_\beta}}} + \sigma^2_{P_{\ket{H_\alpha V_\beta}}} + \sigma^2_{P_{\ket{H_\alpha H_\beta}}}}.
\end{split}
\label{Sigma_E}
\end{equation}

\noindent Finally, for both functions $S$ and $S'$, defined in Eq.~(\ref{CHSH_S}) and Eq.~(\ref{CHSH_S'}), we have the same associated error,
\begin{equation}
    \sigma_S = \sigma_{S'} =
    \sqrt{\sigma^2_{E(\alpha, \beta)} + \sigma^2_{E(\alpha, \beta')} + \sigma^2_{E(\alpha', \beta)} + \sigma^2_{E(\alpha', \beta')}}\,.
\end{equation}

\noindent Also, we define the coverage ratio as the distance in standard deviations between the absolute value of our value obtained in the CHSH inequality and the maximum value that can be obtained clasically, i.e. 2. Thus, 
\begin{equation}
\begin{split}
    \rm{Coverage~Ratio} &= \frac{|\langle S \rangle| - 2}{\sigma_S} ~~~ \rm{ for } ~\ket{\Phi^{+}} ~ \rm{and} ~\ket{\Psi^{-}} \\
    &= \frac{|\langle S' \rangle| - 2}{\sigma_{S'}} ~~~ \rm{ for } ~\ket{\Phi^{-}} ~ \rm{and} ~\ket{\Psi^{+}} \,.
\end{split}
\end{equation}

\end{widetext}

\end{document}